\documentclass[oldversion]{aa}  
\usepackage{graphicx,amsmath}
\usepackage{amssymb}
\usepackage{color}
\usepackage{natbib}             
\usepackage{url}
\usepackage{multirow}
\usepackage{threeparttable}   
\usepackage{soul,mathabx}
\bibpunct{(}{)}{;}{a}{}{,} 

\def\approxinf{%
  \def\p{%
    \setbox0=\vbox{\hbox{$<$}}%
    \ht0=0.6ex \box0 }%
  \def\s{%
    \vbox{\hbox{$\sim$}}%
  }%
  \mathrel{\raisebox{0.7ex}{%
      \mbox{$\underset{\s}{\p}$}%
    }}%
}

\begin{document}

   \title{The Rossiter-McLaughlin effect Revolutions: An ultra-short period planet and a warm mini-Neptune on perpendicular orbits\thanks{Based in part on Guaranteed Time Observations collected at the European Southern Observatory under ESO programme 1104.C-0350(D) by the ESPRESSO Consortium}}                 
        
   \author{    
V.~Bourrier\inst{1},
C.~Lovis\inst{1},
M.~Cretignier\inst{1},   
R.~Allart\inst{2,1},
X.~Dumusque\inst{1},
J.-B.~Delisle\inst{1},
A.~Deline\inst{1}, 
S.~G.~Sousa\inst{7},   
V.~Adibekyan\inst{7,10},
Y.~Alibert\inst{3},
S.~C.~C.~Barros\inst{7,10},
F.~Borsa\inst{12},
S.~Cristiani\inst{4},
O.~Demangeon\inst{7,10},
D.~Ehrenreich\inst{1}, 
P.~Figueira\inst{8,7},   
J.I.~Gonz\'alez~Hern\'andez\inst{5,6},  
M.~Lendl\inst{1},        
J.~Lillo-Box\inst{19},
G.~Lo~Curto\inst{9},  
P.~Di~Marcantonio\inst{4},
C.J.A.P.~Martins\inst{7,8},
D.~M\'egevand\inst{1},
A.~Mehner\inst{9},
G.~Micela\inst{15},
P.~Molaro\inst{4,11},
M.~Oshagh\inst{5,6},
E.~Palle\inst{5,6},
F.~Pepe\inst{1},
E.~Poretti\inst{12},
R.~Rebolo\inst{17,6,18},
N.~C.~Santos\inst{7,10},
G.~Scandariato\inst{13},   
J.~V.~Seidel\inst{1},
A.~Sozzetti\inst{14},
A. Su\'{a}rez Mascare\~{n}o\inst{17,6},
M.~R~Zapatero~Osorio\inst{16},      
        }

\authorrunning{V.~Bourrier et al.}
\titlerunning{The Rossiter-McLaughlin effect Revolutions}

\offprints{V.B. (\email{vincent.bourrier@unige.ch})}

\institute{
Observatoire Astronomique de l'Universit\'e de Gen\`eve, Chemin Pegasi 51b, CH-1290 Versoix, Switzerland
\and 
Department of Physics, and Institute for Research on Exoplanets,Universit\'e de Montr\'eal, Montr\'eal, H3T 1J4, Canada
\and 
Physics Institute of University of Bern, Gesellschaftsstrasse 6, CH-3012 Bern, Switzerland
\and 
INAF, Osservatorio Astronomico di Trieste, via G. B. Tiepolo 11, I-34143, Trieste, Italy
\and 
Instituto de Astrofisica de Canarias, Via Lactea, E-38200 La Laguna, Tenerife, Spain
\and 
Universidad de La Laguna, Departamento de Astrof\'isica, E- 38206, La Laguna, Tenerife, Spain
\and 
Instituto de Astrof\'{\i}sica e Ci\^encias do Espa\c co, CAUP, Universidade do Porto, Rua das Estrelas, 4150-762, Porto, Portugal
\and 
Centro de Astrof\'{\i}sica da Universidade do Porto, Rua das Estrelas, 4150-762 Porto, Portugal
\and 
European Southern Observatory, Karl-Schwarzschild-Str. 2, 85748 Garching bei M\"unchen, Germany
\and 
Departamento de F\'{\i}sica e Astronomia, Faculdade de Ci\^encias, Universidade do Porto, Rua do Campo Alegre, 4169-007 Porto, Portugal
\and 
Institute for Fundamental Physics of the Universe, Via Beirut 2, 34151 Miramare, Trieste, Italy
\and 
INAF, Osservatorio Astronomico di Brera, Via Bianchi 46, 23807 Merate, Italy
\and 
INAF, Osservatorio Astrofisico di Catania, Via S. Sofia 78, 95123 Catania, Italy
\and 
INAF, Osservatorio Astrofisico di Torino, Via Osservatorio 20, 10025 Pino Torinese, Italy
\and 
INAF, Osservatorio Astronomico di Palermo, Piazza del Parlamento 1, 90134 Palermo, Italy
\and 
Centro de Astrobiolog\'ia (CSIC-INTA), Carretera de Ajalvir km 4, 28850 Torrej\'on de Ardoz, Madrid, Spain
\and 
Instituto de Astrof\'{i}sica de Canarias, E-38205 La Laguna, Tenerife, Spain
\and  
Consejo Superior de Investigaciones Cient\'ificas, E-28006 Madrid, Spain
\and 
Centro de Astrobiolog\'{i}a (CAB, CSIC-INTA), Depto. de Astrof\'{i}sica, ESAC campus 28692 Villanueva de la Ca{\~n}ada (Madrid)
}

   \date{} 
 
  \abstract
{Comparisons of the alignment of exoplanets with a common host star and each other can be used to distinguish among concurrent evolution scenarios for the star and the planets. However, multi-planet systems usually host mini-Neptunes and super-Earths, whose sizes make orbital architecture measurements challenging. We introduce the Rossiter-McLaughlin effect Revolutions (RMR) technique, which can access the spin-orbit angle of small exoplanets by exploiting the full extent of information contained in spectral transit time series. We validated the technique through its application to published HARPS-N data of the mini-Neptune HD\,3167c ($P$ = 29.8\,d), refining its high sky-projected spin-orbit angle (-108.9$\stackrel{+5.4}{_{-5.5}}^{\circ}$), and we applied it to new ESPRESSO observations of the super-Earth HD\,3167\,b ($P$ = 0.96\,d), revealing an aligned orbit (-6.6$\stackrel{+6.6}{_{-7.9}}^{\circ}$). Surprisingly different variations in the contrast of the stellar lines occulted by the two planets can be reconciled by assuming a latitudinal dependence of the stellar line shape. In this scenario, a joint fit to both datasets constrains the inclination of the star (111.6$\stackrel{+3.1}{_{-3.3}}^{\circ}$) and the 3D spin-orbit angles of HD\,3167b (29.5$\stackrel{+7.2}{_{-9.4}}^{\circ}$) and HD\,3167c (107.7$\stackrel{+5.1}{_{-4.9}}^{\circ}$). The projected spin-orbit angles do not depend on the model for the line contrast variations, and so, with a mutual inclination of 102.3$\stackrel{+7.4}{_{-8.0}}^{\circ}$, we can conclude that the two planets are on perpendicular orbits. This could be explained by HD\,3167b being strongly coupled to the star and retaining its primordial alignment, whereas HD\,3167c would have been brought to a nearly polar orbit via secular gravitational interactions with an outer companion. Follow-up observations of the system and simulations of its dynamical evolution are required to search for this companion and explore the likelihood of this scenario. HD\,3167\,b (R$_{\rm}$ = 1.7\,R$_{\rm Earth}$) is the smallest exoplanet with a confirmed spectroscopic Rossiter-McLaughlin signal. The RMR technique opens the way to determining the orbital architectures of the super-Earth and Earth-sized planet populations.}

\keywords{}

   \maketitle


\section{Introduction}

Measurements of star-planet alignments are essential to improving our understanding of exoplanet formation and evolution, especially for planets that migrated close to their star. Disk-driven migration is expected to conserve the initial alignment between the angular momentums of the protoplanetary disk and of the planetary orbits (e.g., \citealt{Winn2015}). A variety of formation and migration pathways can, however, lead to misalignments between the spins of a star and of its planets' orbits, namely: a primordial tilt of the star or the protoplanetary disk (e.g., \citealt{Lai2011}, \citealt{Bate2010}, \citealt{Drummond2015}); a tidal torque on the protoplanetary disk from a neighboring star (e.g., \citealt{Lai2014}, \citealt{Batygin2012}; \citealt{Zanazzi2018}); a tidal torque on the inner planetary system from an outer companion (e.g., \citealt{huber2013}); and scattering or secular interactions between the inner planets and an outer planetary or stellar companion (e.g., \citealt{Wu2003}, \citealt{Chatterjee2008}, \citealt{Fabrycky2007}, \citealt{Teyssandier2013}). In this context, determining the orbital architecture of multi-planet systems is of great relevance as the mutual inclinations between the planets, and with the star, can help distinguish among various formation and migration scenarios. Early measurements revealed coplanar orbits that are well aligned with the stellar equator in multi-planet systems (e.g. \citealt{Figueira2012}; \citealt{Hirano2012}; \citealt{SanchisOjeda2012}; \citealt{albrecht2013}, \citealt{Chaplin2013}, \citealt{VanEylen2014}), standing in contrast to the broad distribution of misalignments observed for hot Jupiters (e.g., \citealt{Naoz2012}, \citealt{Albrecht2012}, \citealt{Davies2014}) and suggesting that their orbital planes still trace a primordial alignment with the protoplanetary disk. Interestingly, spin-orbit measurements in young systems ($<$100\,Myr) have shown prograde and aligned orbits (\citealt{Palle2020b}; \citealt{Zhou2020,Mann2020}). Other studies have unveiled substantial misalignments in multi-planet systems (\citealt{huber2013}, \citealt{Walkowicz2013}; \citealt{Hirano2014}, \citealt{Hjorth2021}, \citealt{Zhang2021}), hinting at more complex formation processes (e.g., \citealt{Spalding2014, Spalding2020}) or dynamical histories (e.g. \citealt{Gratia2017}). \\

Analyses of an exoplanet transit at high temporal cadence and with high spectral resolution can be used to resolve the path of the planet across the star and thus measure its obliquity, defined as the angle between the normal to the planetary orbit and the rotation axis of the star (usually only known through its projection $\lambda$ in the plane of the sky). The occultation of the rotating stellar photosphere by the planet creates a signature in the disk-integrated lines of the star, whose spectral position is linked with the spatial position of the occulted stellar regions (the so-called Rossiter-McLaughlin, or RM, effect; \citealt{rossiter1924}; \citealt{mclaughlin1924}). This distortion of the stellar lines results in an anomalous deviation of their RV centroid from the Keplerian curve, whose analysis has traditionally been used to study the RM effect and derive $\lambda$ (see review by \citealt{Triaud2018}). Alternative techniques consist of directly fitting  the disk-integrated lines with a spectral model accounting for the signature of the planet (Doppler tomography; e.g., \citealt{cameron2010a}; \citealt{bourrier2015_koi12}; \citealt{Temple2019}) or retrieving the stellar line occulted by the planet and measuring  its local RV centroid directly (Reloaded RM effect; e.g., \citealt{Cegla2016a}; \citealt{Bourrier2020_HEARTSIII}, \citealt{Kunovac2021}). By using the planet as a probe of the stellar surface, the latter technique provides direct access to its properties and reduces biases in the derived obliquity (\citealt{Cegla2016a}, \citealt{Bourrier2017_WASP8}).\\

In the current paper, we investigate the orbital architecture of the HD\,3167 system, host to three known planets (Sect.~\ref{sec:sys_prop}). We used high-resolution optical spectroscopy to analyze the Rossiter-McLaughlin effect of HD\,3167b and c (Sect.~\ref{sec:obs_red}), using a new technique that we describe and apply in Sect.~\ref{sec:RMR}. We discuss the orbital architecture and possible history of the HD\,3167 system in light of our results in Sect.~\ref{sec:orb_arch} and we present our conclusions in Sect.~\ref{sec:conclu}.


\section{HD\,3167 system}
\label{sec:sys_prop}

HD\,3167 is a nearby (47\,pc) and bright (V = 9) K0-type star hosting three known planets (\citealt{Vanderburg2016}, \citealt{Gandolfi2017}, \citealt{Christiansen2017}). The innermost planet, HD\,3167b, is a super-Earth ($R_{p}$ = 1.70$\stackrel{+0.18}{_{-0.15}}$\,$R_{\Earth}$, $M_{p}$ = 5.02$\pm$0.38\,$M_{\Earth}$) on an ultra-short period of 0.96\,d and the outermost one, HD\,3167c, is a mini-Neptune ($R_{p}$ = 3.01$\stackrel{+0.42}{_{-0.28}}$\,$R_{\Earth}$, $M_{p}$ = 9.80$\stackrel{+1.30}{_{-1.24}}$\,$M_{\Earth}$) on a 29.84\,d orbit. While HD\,3167b and HD\,3167c transit, this is not the case for HD\,3167d ($M_{p}\,sin\,i  = $6.90$\pm$0.71\,$M_{\Earth}$), which orbits in between them on a 8.51\,d orbit. HD\,3167c is on a highly misaligned orbit around the star and dynamical analysis has shown that its orbit is likely nearly coplanar with that of HD\,3167d, albeit with a mutual inclination, likely ranging between 2.3 and 21$^{\circ}$ (\citealt{Dalal2019}). \\

Properties of the HD\,3167 system useful to our analysis were set to the values derived by \citealt{Christiansen2017} (hereafter, C17). Precise values for the ephemeris, however, are critical to the analysis of RM signals (e.g., \citealt{CasasayasBarris2021}); C17's ephemeris yield 1$\sigma$ uncertainties of 19\,min for HD\,3167b and 16\,min for HD\,3167c at the time of our observations. As is common for close-in planets, these uncertainties are dominated by the error propagated from the period, which increases linearly with the number of orbital revolutions. To address this issue, we exploited mid-transit times derived from recent transit observations of HD\,3167b with the CHEOPS satellite ($T_\mathrm{0}$ = 2459141.90513$\stackrel{+0.00048}{_{-0.00050}}$; Bourrier et al. in prep.) and of HD\,3167c with the Hubble Space Telescope (\citealt{Guilluy2021}). The large number of planetary transits elapsed between the mid-transit time $T_\mathrm{0}^\mathrm{C17}$ derived by C17 and those recent $T_\mathrm{0}^\mathrm{new}$ measurements allow for a substantial refinement of the ephemeris as:\\

\begin{equation}
\begin{split}
P=&(T_\mathrm{0}^\mathrm{new}-T_\mathrm{0}^\mathrm{C17})/n \\
\sigma(P) =& \sqrt{\sigma(T_\mathrm{0}^\mathrm{new})^2 + \sigma(T_\mathrm{0}^\mathrm{C17})^2}/n \\
T_\mathrm{0}^\mathrm{RM} =& (n_\mathrm{new} T_\mathrm{0}^\mathrm{C17} + n_\mathrm{C17} T_\mathrm{0}^\mathrm{new})/n  \\
\sigma(T_\mathrm{0}^\mathrm{RM}) =& \sqrt{\left(n_\mathrm{new} \sigma(T_\mathrm{0}^\mathrm{C17})\right)^2 + \left(n_\mathrm{C17} \sigma(T_\mathrm{0}^\mathrm{new})\right)^2}/n      
\end{split}
\label{eq:ephem}  
,\end{equation}
where $P$ indicates the orbital period, $\sigma$ the uncertainty on the considered parameter (C17 and new parameters are assumed to be uncorrelated), and $n = n_\mathrm{new}+n_\mathrm{C17}$ (with $n_\mathrm{C17}$ and $n_\mathrm{new}$ the number of orbital revolutions elapsed between a given RM observation and the C17 and new mid-transit times, respectively). We derive $P_\mathrm{b}$ = 0.9596550$\pm$6.0$\times$10$^{-7}$\,d and $P_\mathrm{c}$ = 29.846449$\pm$1.9$\times$10$^{-5}$\,d and we decrease to below 1\,min the uncertainties on the mid-transit time at the epochs of the HARPS-N ($T_\mathrm{0}^\mathrm{RM}$ = 2457663.59662$\pm$4.4$\times$10$^{-4}$\,BJD) and ESPRESSO ($T_\mathrm{0}^\mathrm{RM}$ = 2458534.44470$\pm$6.7$\times$10$^{-4}$\,BJD) observations. \\

\section{Observations and data reduction}
\label{sec:obs_red}

\subsection{Transit observations}
\label{sec:obs}

We exploited two datasets on the HD\,3167 system, one obtained with ESPRESSO during the transit of HD\,3167\,b and the other obtained with HARPS-N during the transit of HD\,3167\,c. Observing conditions for the two transits are given in Table~\ref{observation W127}. Observations outside of the transit are essential for analyses of the RM effect: to measure the stable stellar flux, to determine the precise RV zero point at the epoch of the observations, and to search for variations due to the Earth's atmosphere and the instrument, which can then be corrected over the entire data set. About half of the exposures were obtained during transit for both HD\,3167\,b and HD\,3167\,c, with out-of-transit exposures obtained both before and after the transits (e.g., Fig.~\ref{fig:RawProp_Compa}). \\

ESPRESSO (\citealt{Pepe2021}) is a fiber-fed, ultra-stabilized high-resolution echelle spectrograph installed at the Very Large Telescope (VLT) at ESO's Paranal site. The light is dispersed on 85 spectral orders (each order being covered by two independent slices on the detectors) from 380 to 788\,nm. HD\,3167 was observed in the frame of Guaranteed Time Observations (Prog: 1104.C-0350(D)) using one of the four 8 m Unit Telescopes (UT3) of the VLT. A constant exposure time was maintained throughout the night, yielding 39 exposures with an average signal-to-noise ratio (S/N) per pixel of 50 at 550\,nm (Table~\ref{observation W127}). We used fiber A to observe the target and fiber B to monitor the sky. The sky signature was not corrected for, as no detrimental contamination was identified. Spectra were extracted from the detector frames, corrected for, and calibrated by version 2.2.8 of the Data Reduction Software (DRS) pipeline. They were then passed through weighted cross-correlation (\citealt{baranne1996}; \citealt{pepe2002}) with a custom-built numerical mask (Sect.~\ref{sec:CCF_masks}), yielding cross-correlation functions (CCFs) with a step corresponding to the ESPRESSO pixel width (500\,m\,s$^{-1}$). Observations were obtained using the HR21 mode (R$\sim$140 000), which yields an instrumental resolution of $\sim$2.1\,km\,s$^{-1}$. \\

HARPS-N (\citealt{Cosentino2012}) is a fiber-fed, cross-dispersed, environmentally stabilized echelle spectrograph installed at the 3.58 m Telescopio Nazionale Galileo (TNG, La Palma, Spain). The light is dispersed on 69 spectral orders from 383 to 690\,nm. HD\,3167 was observed in the frame of a Director's Discretionary Time program (PI: Hebrard), which was used to study the RM effect of HD\,3167\,c (\citealt{Dalal2019}). Here we use these archival data to validate our new technique and refine the spin-orbit angle of HD\,3167\,c. Following \citet{Dalal2019}, we discarded the second night of the program obtained in poor weather conditions and with little out-of-transit baseline. Our analysis was performed on the 35 first exposures in the first night, obtained with an exposure time of 900\,s and an average S/N of 87 at 550\,nm (Table~\ref{observation W127}). The last exposure was discarded because of its low exposure time (427\,s), high airmass (4.2), and low resulting S/N (33). The remaining four last exposures still have airmass larger than 2, but their S/N remains high, they only contribute to the reference spectrum for the star (Sect.~\ref{sec:step1}), and they do not show any spurious features (Fig.~\ref{fig:Res_map_HD3167c}). Fiber A was set on the star and fiber B on a thorium-argon lamp to ensure a higher RV precision. The data were extracted and CCFs calculated using version 2.2.8 of the ESPRESSO pipeline adapted to HARPS-N (\citealt{Dumusque2021}), using a custom-built mask (Sect.~\ref{sec:CCF_masks}) and a step corresponding to the pixel width (820\,m\,s$^{-1}$). HARPS-N provides a resolving power R$\sim$115000, corresponding to an instrumental resolution of $\sim$2.6\,km\,s$^{-1}$. \\

The variability of the extinction caused by Earth’s atmosphere modifies the spectral flux balance of the stellar spectra in a different way throughout each night. In turn, this so-called color effect impacts the shape of the CCFs and can bias the measurements derived from transit time series (e.g., \citealt{bourrier2014b}, \citealt{Bourrier_2018_Nat}, \citealt{Wehbe2020}). To avoid these biases, the flux balance of HD\,3167 spectra was reset to a stellar spectrum template representative of its spectral type by the ESPRESSO and HARPS-N DRS.

\begin{table}[h]
\centering
\caption{Observational log.}
\begin{tabular}{lcccc}
\hline
Observing night & 2016-10-01 & 2019-02-20\\
\hline
Transit  &  HD\,3167\,c    &  HD\,3167\,b  \\
Instrument & HARPS-N & ESPRESSO \\
Spectra & 35 & 39 \\
In-transit & 20 & 16 \\
Pre/post-transit & 10/5 & 8/15 \\
 t$_{\rm exp}$ [s] & 900 & 300 \\
 Airmass & 1.10 - 3.8 & 1.14 - 1.86 \\
 S/N@550\,nm & 56 - 115 & 34 - 68 \\
\hline
\end{tabular}
\begin{tablenotes}[para,flushleft]
\end{tablenotes}
\label{observation W127}
\end{table}


\subsection{Improved CCF masks}
\label{sec:CCF_masks}

\begin{figure*}
\begin{minipage}[h!]{\textwidth}
\includegraphics[trim=0cm 0cm 0cm 0cm,clip=true,width=\columnwidth]{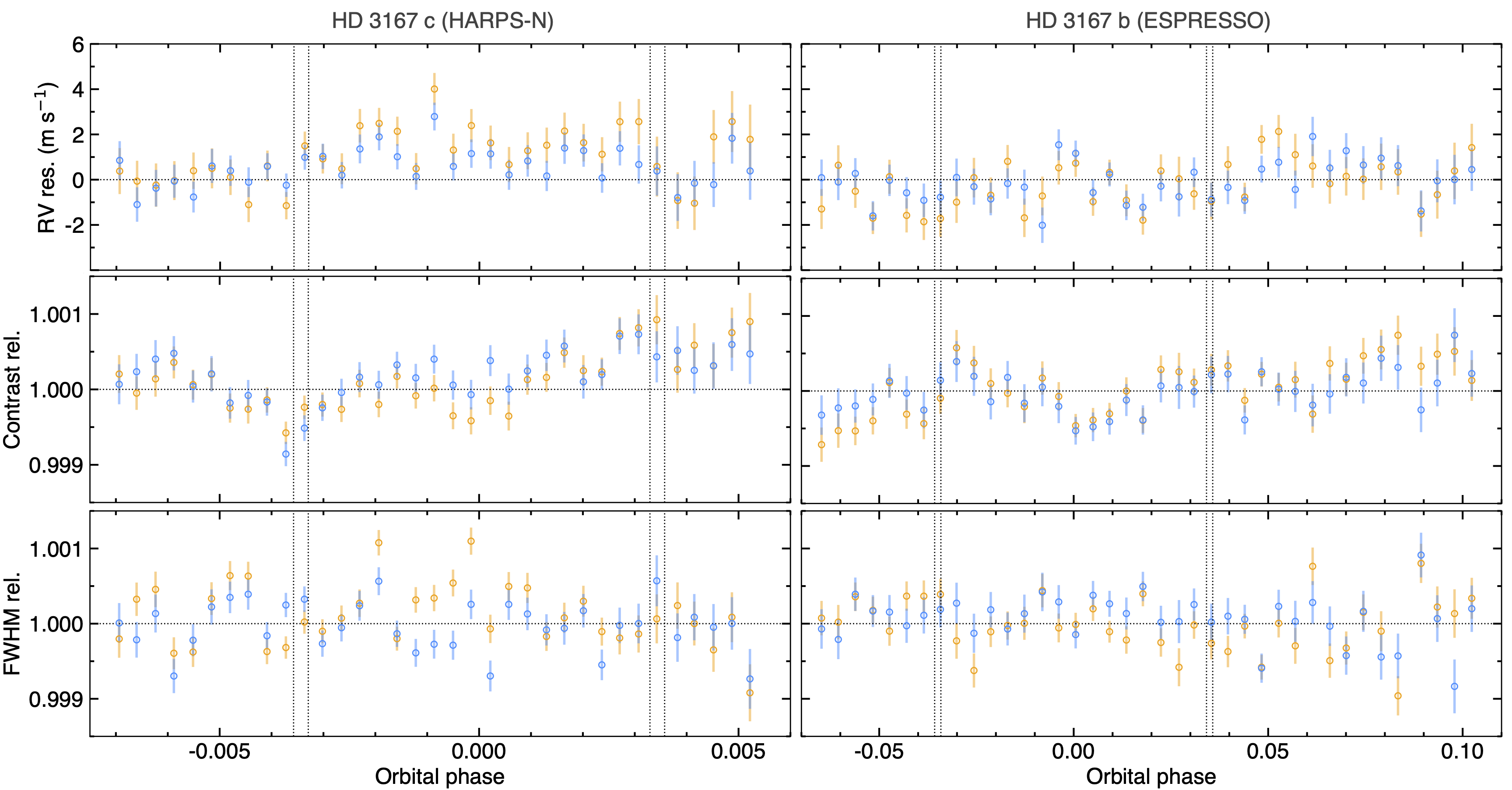}
\centering
\end{minipage}
\caption[]{Time series properties derived from the HARPS-N (left panel) and ESPRESSO (right panel) CCF$_\mathrm{DI}$ calculated with the masks previously used by the instruments' DRS (orange) and with masks built with an improved definition of the line weights and a line selection specific to HD\,3167 (blue). Top panels show Keplerian RV residuals. Middle and bottom panels show the contrast and FWHM relative to their out-of-transit mean value, so that they can be compared between the different masks. Vertical dashed lines show the transit contacts of the observed HD\,3167 planets.}
\label{fig:RawProp_Compa}
\end{figure*}

CCFs calculated from the observed spectra correspond to the stellar lines averaged over the disk of the star (hereafter CCF$_\mathrm{DI}$, for disk-integrated CCFs). Each CCF$_\mathrm{DI}$ is automatically fitted by the DRS with a Gaussian model, yielding time series of their contrast, full-width at half-maximum (FWHM), and radial velocity centroid (RV). After subtracting a Keplerian model calculated with the properties of HD\,3167\,b, \,c, and \,d from C17 from the RVs, we assessed the quality and stability of the measurements via these three properties (Fig.~\ref{fig:RawProp_Compa}). In-transit CCF$_\mathrm{DI}$ were excluded from this analysis since they show anomalous deviations due to the planets occulting local regions of the stellar disk.

For both HARPS-N and ESPRESSO, the dispersion of the out-of-transit CCF$_\mathrm{DI}$ properties was found to be larger than their averaged photon-noise uncertainty, suggesting the presence of systematic variations (Table~\ref{tab:disp}). Order-by-order and line-by-line analyses (\citealt{Dumusque2018}) revealed such increased dispersions in the bluest orders and in specific orders over the green band. Short wavelengths exhibit many blended stellar lines, which are more sensitive to activity (\citealt{Anglada2012}). At longer wavelengths, several orders contain stellar lines with deep narrow cores and wide pressure-broadened wings, such as the sodium doublet, whose profiles may not be well captured by a Gaussian model, thus decreasing the stability of the CCF$_\mathrm{DI}$ they contribute to. We hypothesized that the current DRS masks were amplifying the contribution of these variable lines to the dispersion of the CCF$_\mathrm{DI}$ properties by overweighting them. Indeed, the weight of a mask line is defined as the square of its depth, which corresponds to the line RV precision assuming that all lines have the same FWHM (\citealt{Lovis2010}). In the case of an inaccurate determination of the continuum level for blended lines or lines with broad wings, the weight in the mask will not correctly reflect  the RV precision of the lines.

We therefore developed new masks with weights more representative of the photonic error on the line positions, and with a line selection optimized for the present observations (as in \citealt{Bourrier2020_HEARTSIII}). Out-of-transit spectra were first normalized with the RASSINE code (\citealt{Cretignier2020b}) and aligned to a common rest frame using their individual RV measurements. Master spectra were then obtained by taking the median of the spectra time series, for the HARPS-N and ESPRESSO data independently. Because upper envelop algorithms such as the ``Jarvis March'' used in RASSINE can yield inaccurate continuum levels, the master continuum was corrected for using a stellar template from the POLLUX database (\citealt{Palacios2020}), selected for a star with similar effective temperature and gravity as HD\,3167. Such an absolute correction of the continuum level is described in the appendix of \citealt{Cretignier2021} and can be understood as a matching of the master spectra and stellar template upper envelopes. More accurate line depths can be calculated with this corrected continuum, and we rejected shallow lines with contrast smaller than 10\%.

Line positions, identified by local minima surrounded by two local maxima, were fitted with a second order polynomial function of the core of spectral lines (\citealt{Cretignier2020a}). Weights on individual lines were defined using the formula from \citet{Bouchy2001}, integrating the line profiles between $\pm$5\,km\,s$^{-1}$ around the fitted line center. We used a standard telluric template (\citealt{Smette2015}, \citealt{Cretignier2020a}) to identify and reject stellar lines whose relative depth with contaminant telluric lines is larger than 2\%. Optimized line selection indeed requires a trade-off (sometimes referred to as a noise-bias trade-off) between rejecting lines affected by systematics and keeping as many lines as possible to increase the S/N. Telluric contamination was computed for the barycentric Earth radial velocity (BERV) span of the observations, $\pm$3\,km\,s$^{-1}$ to account for the width of the tellurics lines. This selection may thus slightly depend on the observing epoch, but this approach allows excluding the smallest possible range of telluric lines contaminating the data on the observed nights, rather than a wider range accounting for all possible BERVs, as in generic masks. \\

A second cutoff was applied to reject lines unsuitable for the determination of precise RV measurements. We first measured the line-by-line RVs in each exposure (\citealt{Dumusque2018}), and then calculated the weighted rms for each line over the out-of-transit time series. Lines presenting a rms larger than 50\,m\,s$^{-1}$ (for HARPS-N) and 80\,m\,s$^{-1}$ (for ESPRESSO) were rejected, this cutoff again being determined as the best noise-bias trade-off. The larger cutoff for the ESPRESSO data is linked to the larger dispersion of the line-by-line RVs, itself related to the lower S/N of this dataset (Table~\ref{tab:disp}). The underlying assumption behind the cutoff is that out-of-transit residual RVs are expected to be constant on average, the observed dispersion arising mainly from lines affected by stellar activity. This second line selection can thus also differ between instruments on different telescopes, as their sensitivity to bias depend on the photon noise level. We note that using the rms as a metric does not allow us to know whether rejected lines are affected by systematics or are intrinsically less precise, but the gain brought by rejecting lines affected by systematics is higher than the loss of lines with a high level of white noise. \\

\begin{figure*}
\begin{minipage}[h!]{\textwidth}
\includegraphics[trim=0cm 0cm 0cm 0cm,clip=true,width=\columnwidth]{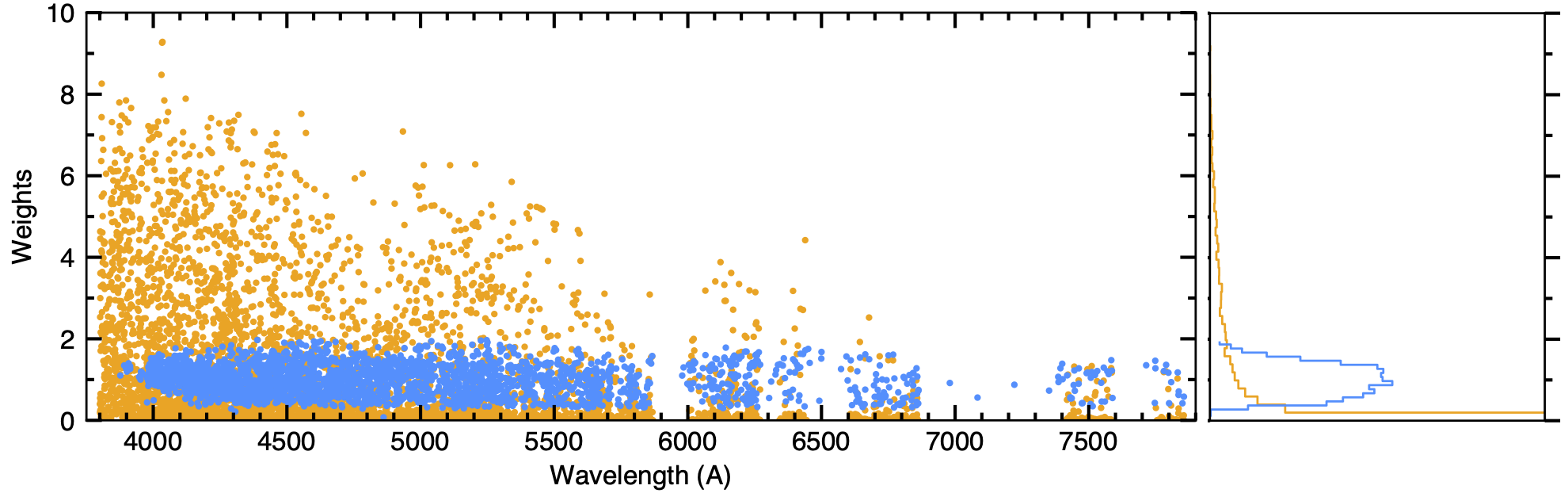}
\centering
\end{minipage}
\caption[]{Weight comparison between the mask currently used in the ESPRESSO DRS for a G9-type star (orange) and the custom mask we built for HD\,3167 (blue). Each point corresponds to one of the masks' lines. Weights are normalized to a mean unity. The right panel shows histograms of the line weights.}
\label{fig:Weights_compa}
\end{figure*}

Figure~\ref{fig:Weights_compa} presents a comparison of the distributions of weights between the generic and custom ESPRESSO masks. The generic mask contains a large number of shallow lines that even when combined together bring more noise than information to the CCF$_\mathrm{DI}$; thus they are excluded from the custom mask. We can also see how the generic mask overweighs a substantial fraction of stellar lines, particularly toward shorter wavelengths. Besides an improved line weighing, the RV dispersion cutoff we applied naturally excludes both the low-precision and strongly active lines at the extreme blue of the spectrum. Together, the tellurics and RV dispersion cutoffs decrease the total weight of the line selection by an average of 20\%, which is equivalent to an increase in the RV uncertainties by 12\%. However, the resulting reduction in bias considerably improves the dispersion of the CCF$_\mathrm{DI}$ constrast, FWHM, and RVs (Fig.~\ref{fig:RawProp_Compa}). The custom masks yield shallower but substantially narrower CCF$_\mathrm{DI}$ (Table~\ref{tab:disp}) because they do not include many blended lines in the blue part of the spectrum, which artificially broaden the FWHM of the CCF. The custom masks further yield much more stable FWHM measurements for both the HARPS-N and ESPRESSO data (possibly because it is a known activity tracer and the new weights reduce the contribution of active lines) and a remarkable stabilisation of the contrast in the ESPRESSO data. The dispersion of all properties is closer to their photonic error, especially for the contrast and RV residuals, with a substantial reduction of the dispersion on the RV residuals for both HARPS-N (by 34\%) and ESPRESSO (by 25\%).\\

\begin{table}[h]
\centering
\caption{Out-of-transit CCF$_\mathrm{DI}$ properties.}
\begin{tabular}{lcccc}
\hline
Instrument   &   \multicolumn{2}{c}{HARPS-N}&   \multicolumn{2}{c}{ESPRESSO}   \\
\hline
Mask & Classic & Custom & Classic & Custom\\
\hline
\hline
$<$Contrast$>$ (\%)      &  65.936  &  56.694    & 66.569 &  55.238  \\
e$_\mathrm{Contrast}$ (\%)       &  0.016 &  0.015    & 0.015  & 0.014  \\
$\sigma_\mathrm{Contrast}$ (\%)  & 0.025   &  0.020    &  0.027 &  0.015  \\
$\sigma_\mathrm{Contrast}^\mathrm{rel}$ (ppm)  &         382  &  360    &  407 &  268\\
\hline
$<$FWHM$>$ (km\,s$^{-1}$)  &  8.1129   &  6.4130    &  7.7244 &   6.1144 \\
e$_\mathrm{FWHM}$ (m\,s$^{-1}$) &  2.0    & 1.7    &  1.8 &  1.6 \\
$\sigma_\mathrm{FWHM}$ (m\,s$^{-1}$) & 3.5  &   2.1   &  3.2 &  2.2 \\
$\sigma_\mathrm{FWHM}^\mathrm{rel}$ (ppm)   &  435  &  322    &  410  & 365\\
\hline
e$_\mathrm{RV}$ (cm\,s$^{-1}$) & 99  &  83    & 88  &  80  \\
$\sigma_\mathrm{RV}$ (cm\,s$^{-1}$)  &  109  &  72    &  111 & 83\\
\hline
\end{tabular}
\begin{tablenotes}[para,flushleft]
Notes: $\sigma$ indicates standard deviations of the contrast, FWHM, and Keplerian RV residuals with respect to their mean out-of-transit value ($<$x$>$). The $\sigma^\mathrm{rel}$ dispersions have been normalized by this mean to allow for a direct comparison between masks. e$_\mathrm{x}$ indicate the mean out-of-transit error on $x$.
\end{tablenotes}
\label{tab:disp}
\end{table}


\section{The Rossiter-McLaughling effect Revolutions}
\label{sec:RMR}

\subsection{Motivation}

In recent years, a new method has proven more efficient at analyzing the Rossiter-McLaughlin effect of exoplanets at higher precision than the classical RM technique (e.g., \citealt{Ohta2005}, \citealt{Gimenez2006}, \citealt{Hirano2011}, \citealt{Boue2013}.). The reloaded RM (RRM) technique (\citealt{Cegla2016}, see also \citealt{Bourrier2017_WASP8,Bourrier_2018_Nat,Bourrier2020_HEARTSIII,Ehrenreich2020,Allart2020,Kunovac2021}) isolates the local CCFs (heareafter CCF$_\mathrm{loc}$) from the regions of the stellar photosphere that are occulted by a planet during its transit. The RVs of the planet-occulted regions can then be derived as the centroids of the average stellar line contained in each CCF$_\mathrm{loc}$. Fitting those RVs with a realistic model (\citealt{Cegla2016,Bourrier2020_HEARTSIII}) yields constraints on the stellar surface velocity field and since they directly trace the path of the planet across the star, on the orbital architecture of the system as well. 

In its current implementation, however, this approach relies on the possibility to detect and fit the average stellar line in an individual CCF$_\mathrm{loc}$ with a simple $\chi^2$ minimization. Often, this is not possible in low-flux CCF$_\mathrm{loc}$ at the limbs of the star, which are darker at optical wavelengths and only partially occulted by the planet (e.g., \citealt{Bourrier_2018_Nat}). More generally, a RRM analysis cannot be carried out if the stellar line is too shallow, the host star too faint, or the transit depth too small. Up to now, this caveat has stood in the way of determining the orbital architecture of the smallest planets transiting all but the brightest stars (\citealt{Kunovac2021}). 

To address these limitations, we present a new approach to the analysis of the RM effect, which we call the RM effect Revolutions (RMR) method. The first step (Sect.~\ref{sec:step1}), up to the extraction of the planet-occulted CCFs, is the same as in the traditionnal RRM approach. The second step (Sect.~\ref{sec:step2}) improves on the analysis of the average stellar line in individual CCF$_\mathrm{loc}$ by using MCMC exploration. The third step (Sect.~\ref{sec:step3}) exploits the full temporal and spectral information contained in the CCF$_\mathrm{loc}$ time series to derive the stellar surface and orbital architecture properties with a higher precision.\\


\subsection{Step 1: Extraction of planet-occulted CCFs}
\label{sec:step1}

\begin{figure}
\includegraphics[trim=0cm 0cm 0cm 0cm,clip=true,width=\columnwidth]{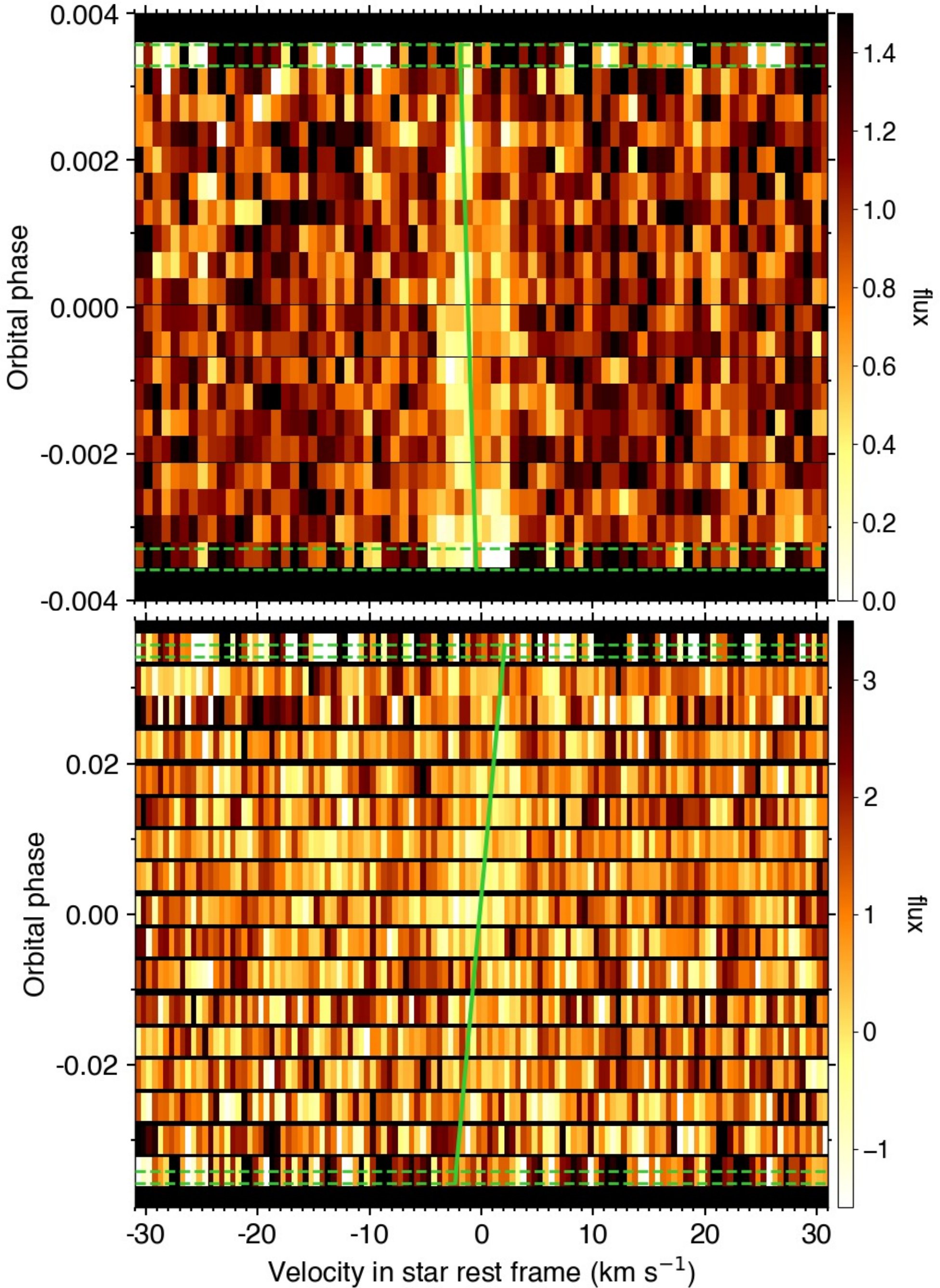}
\centering
\caption[]{Maps of the CCF$_\mathrm{intr}$ during the transits of HD\,3167\,c (HARPS-N, upper panel) and HD\,3167\,b (ESPRESSO, lower panel). Transit contacts are shown as green dashed lines. Values are colored as a function of their normalized flux and plotted as a function of RV in the stellar rest frame (in abscissa) and orbital phase (in ordinate). HD\,3167c is large enough that the stellar line can be distinguished in individual exposures, while the small transit depth of HD\,3167b makes the line undetectable in most exposures even with the VLT/ESPRESSO data. The green solid lines show the best-fit models for the stellar surface RVs derived with the joint RMR fit to both data sets.}
\label{fig:CCFintr_map}
\end{figure}

The CCF$_\mathrm{DI}$ reduced by the DRS are first aligned by correcting for the Keplerian motion of the star induced by all planets in the system. The relative flux level of the CCF$_\mathrm{DI}$ is lost due to extinction by Earth atmosphere (Sect.~\ref{sec:obs}). The continuum of CCF$_\mathrm{DI}$ is thus scaled to reflect the planetary disk absorption during the studied transit, using light curves computed with the batman package (\citealt{Kreidberg2015}). The Keplerian and transit models are calculated using properties from C17 and our refined ephemeris. The CCF$_\mathrm{DI}$ outside of the transit are co-added to build a master-out CCF$_\mathrm{DI}$ representative of the unocculted star. By construction, the masks built for HD\,3167 are defined in the star rest frame (Sect.~\ref{sec:CCF_masks}), but we corrected all CCF$_\mathrm{DI}$ for a small residual offset of 37\,m\,s$^{-1}$ measured via a fit to the masters-out of each dataset. Residual CCF$_\mathrm{loc}$ are obtained by subtracting the scaled CCF$_\mathrm{DI}$ from their corresponding master-out. We note that the new versions of the HARPS-N and ESPRESSO DRS propagate errors from the raw spectra to the CCF$_\mathrm{DI}$, allowing us to further propagate the errors on the CCF$_\mathrm{loc}$ and removing the need to use continuum dispersion as estimate for flux uncertainties (e.g., \citealt{Cegla2016}, \citealt{Bourrier2017_WASP8}). Furthermore, rather than working with CCF$_\mathrm{loc}$ as in previous RRM studies, we reset them to a common flux level by dividing their continuum by the flux scaling applied to the CCF$_\mathrm{DI}$. These ``intrinsic'' CCF$_\mathrm{intr}$, shown in Fig.~\ref{fig:CCFintr_map}, are thus independent of planetary occultation and of the effect of limb-darkening on the continuum flux. They only trace variations in the local stellar line profiles, allowing for a more direct comparison.\\

\subsection{Step 2: Analysis of individual exposures}
\label{sec:step2}

In the traditional RRM approach, each CCF$_\mathrm{loc}$ is fitted individually with a stellar line model via Levenberg-Marquardt least-squares minimization. A single criterion (e.g., \citealt{Bourrier_2018_Nat}) is used to determine whether the local stellar line is considered detected and whether the best-fit properties can subsequently be used for analysis. This approach has several limitations; for example, the detectability of the stellar line and its exploitation depends on an arbitrary threshold calculated using the fit properties. If the data have systematic errors, the $\chi^2$ minimization makes it difficult to check that the model was fitted to the stellar line rather than to a spurious feature. If the properties of the stellar line model do not have Gaussian uncertainty distributions, the $\chi^2$ minimization can then bias their best-fit estimates and errors.

In the RMR approach, we improve on the analysis of individual planet-occulted CCFs by using a Bayesian approach. A model of the stellar line is fitted to the CCF$_\mathrm{intr}$ in each exposure, sampling the posterior distributions of the model parameters using \textit{emcee} MCMC (\citealt{Foreman2013}) and allowing for a variety of priors. In the present study we use a simple Gaussian model with jump parameters set to the contrast, FWHM, and RV centroid, and we chose uninformative priors whose main role is to prevent the MCMC walkers from diverging. A uniform prior distribution between [-5 , 5]\,km\,s$^{-1}$ is set on the RV centroid, considering that stellar surface RVs cannot be larger than the maximum projected stellar rotational velocity (about 2.5\,km\,s$^{-1}$ for HD\,3167, as derived from RM analysis or stellar spectroscopy; \citealt{Dalal2019}). The flux in the core of stellar absorption lines is bound by zero and the local continuum (corresponding to a contrast range of [0 , 1]), but we set a uniform prior distribution on the line contrast over a large enough range ([-2 , 2]) that we can assess the impact of noise and potentially noisy features on the line detection. A uniform prior distribution between [0 , 20]\,km\,s$^{-1}$ is set on the FWHM, about three times larger than the average FWHM of the local stellar line (see below). Walkers are initialized randomly over the prior ranges to ensure a thorough exploration of the parameter space. In the present analysis, we empirically chose  (checking for convergence of the chains) to use 100 walkers running for 2000 steps, with a burn-in phase of 500 steps. We tested the model against the data over [-50 , 50]\,km\,s$^{-1}$ in the star rest frame. The median of the posterior probability distributions (PDF) are considered as the derived values for the model parameters. To better account for multimodal and asymmetric PDFs we define 1$\sigma$ uncertainty ranges using highest density intervals (HDIs), which contains 68.3\% of the PDF mass such that no point outside the interval has a higher density than any point within it.


\subsubsection{Single-exposure analysis for HD\,3167c}
\label{sec:step2_c}

\begin{figure*}
\begin{minipage}[h!]{\textwidth}
\includegraphics[trim=0cm 0cm 0cm 0cm,clip=true,width=\columnwidth]{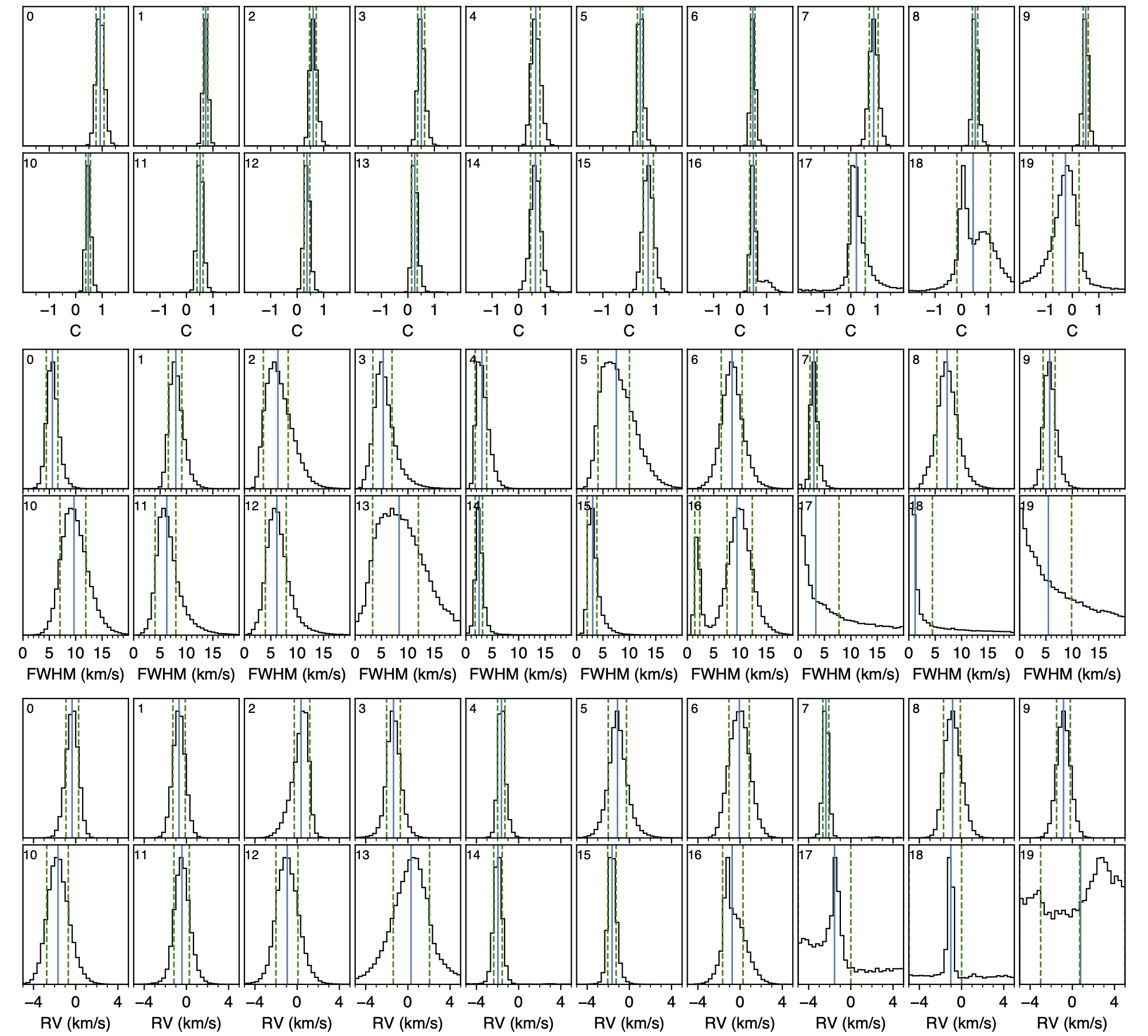}
\centering
\end{minipage}
\caption[]{PDFs of the contrast (upper panels), FWHM (middle panels), and RV centroids (lower panels) of the Gaussian line model fitted to individual CCF$_\mathrm{intr}$ during the transit of HD\,3167c. Deep blue lines indicate the PDF median values, with dashed green lines showing the 1$\sigma$ highest density intervals. In-transit exposure indexes are shown in each subplot.}
\label{fig:PDF_HD3167c_indiv}
\end{figure*}

Results from the MCMC exploration of the HD\,3167c transit are displayed in Fig.~\ref{fig:PDF_HD3167c_indiv}. Most exposures show well-defined PDFs, narrow enough that they constrain the model parameters within the prior ranges. In such cases, the local stellar line can be detected in individual CCF$_\mathrm{intr}$ via $\chi^2$ minimization, as in the RRM study performed by \citet{Dalal2019}. Nonetheless, several exposures show PDFs with asymmetrical profiles, highlighting the interest of a Bayesian approach and the combination of median + HDI to interpret the fit results and prevent biases. Furthermore, this approach reveals that the last three exposures have a contrast and FWHM PDFs that are consistent with null values, which shows that the flux noise in these CCF$_\mathrm{intr}$ is too large for the stellar line to be detected. The S/N of the spectra decreased over the transit as airmass increased, dropping below the average S/N in the night for the last three in-transit exposures. Hereafter, we thus exclude these exposures from all analysis, as they bring more noise than constraints to the interpretation of the transit time series.\\

\begin{figure}
\includegraphics[trim=0cm 0cm 0cm 0cm,clip=true,width=\columnwidth]{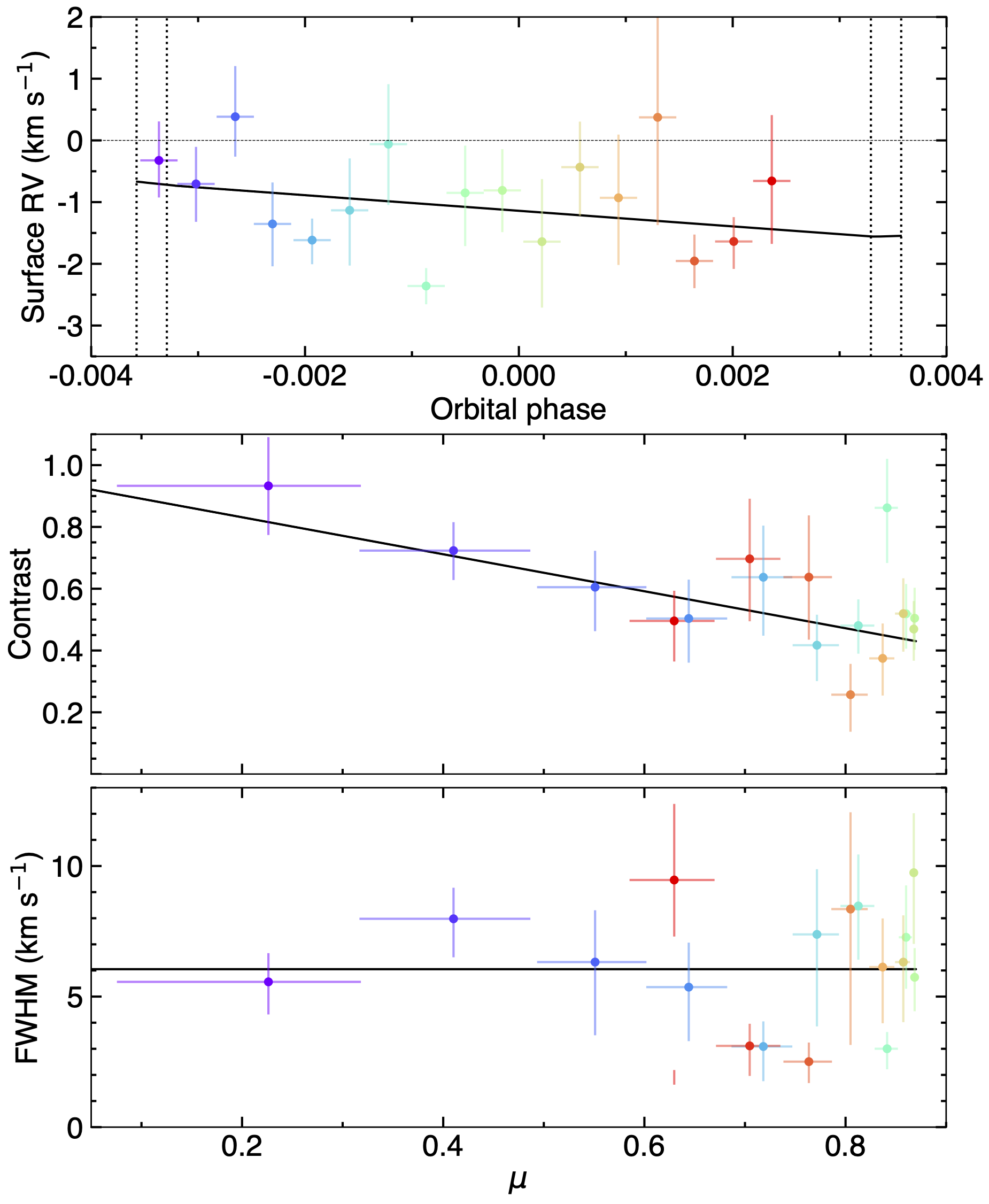}
\centering
\caption[]{Properties of the stellar surface regions occulted by HD\,3167\,c. Data points are colored as a function of orbital phase. Horizontal bars indicate the range of abscissa property covered by the planet in each exposure. Vertical bars indicate the 1$\sigma$ HDI (multiple in case of multimodal PDFs). Solid black curves are the best-fit models to each property, derived using the RMR approach. \textit{Top panel:} Surface RVs as a function of orbital phase. Dashed vertical lines are the transit contacts. The horizontal dashed line indicates the stellar rest velocity. \textit{Bottom panel:} Local contrast and FWHM as a function of $\mu$. Model values correspond to the observed line, i.e., they account for the instrumental convolution.}
\label{fig:PropLoc_HD3167c}
\end{figure}

Local RVs from individual exposures have small enough uncertainties that they clearly show the signature of the RM effect from HD\,3167c, which allowed \citet{Dalal2019} to fit them using the traditional RRM approach (Fig.~\ref{fig:PropLoc_HD3167c}). The local RV series is negative and slightly decreases with phase, showing that HD\,3167c crosses the stellar hemisphere rotating toward us, and moves farther away from the projected stellar spin axis over the course of its transit. This configuration yields the redshifted anomaly visible in the RV centroids of the CCF$_\mathrm{DI}$ (Fig.~\ref{fig:RawProp_Compa}). An analysis of the derived properties in Fig.~\ref{fig:PropLoc_HD3167c} suggests possible center-to-limb variations in local contrast (see e.g., \citealt{Cavallini1985}, \citealt{Cavallini1985,Lohner2019} for the Sun), while uncertainties on the local FWHM are too large to draw any conclusions on center-to-limb variations.

\subsubsection{Single-exposure analysis for HD\,3167b}
\label{sec:step2_b}

\begin{figure*}
\begin{minipage}[h!]{\textwidth}
\includegraphics[trim=0cm 0cm 0cm 0cm,clip=true,width=\columnwidth]{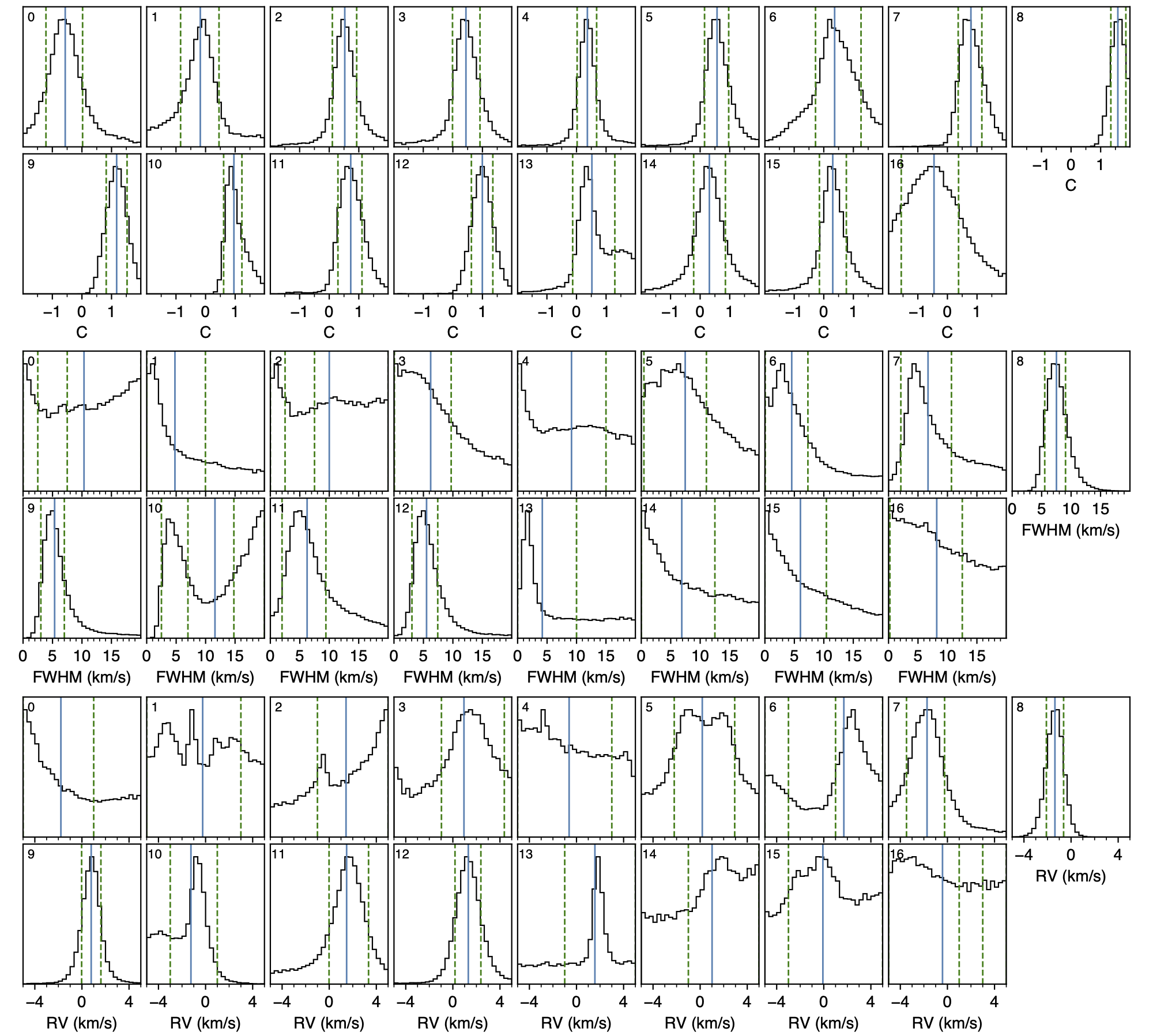}
\centering
\end{minipage}
\caption[]{Same as Fig.~\ref{fig:PDF_HD3167c_indiv} for the HD\,3167b ESPRESSO transit.}
\label{fig:PDF_HD3167b_indiv}
\end{figure*}

\begin{figure}
\includegraphics[trim=0cm 0cm 0cm 0cm,clip=true,width=\columnwidth]{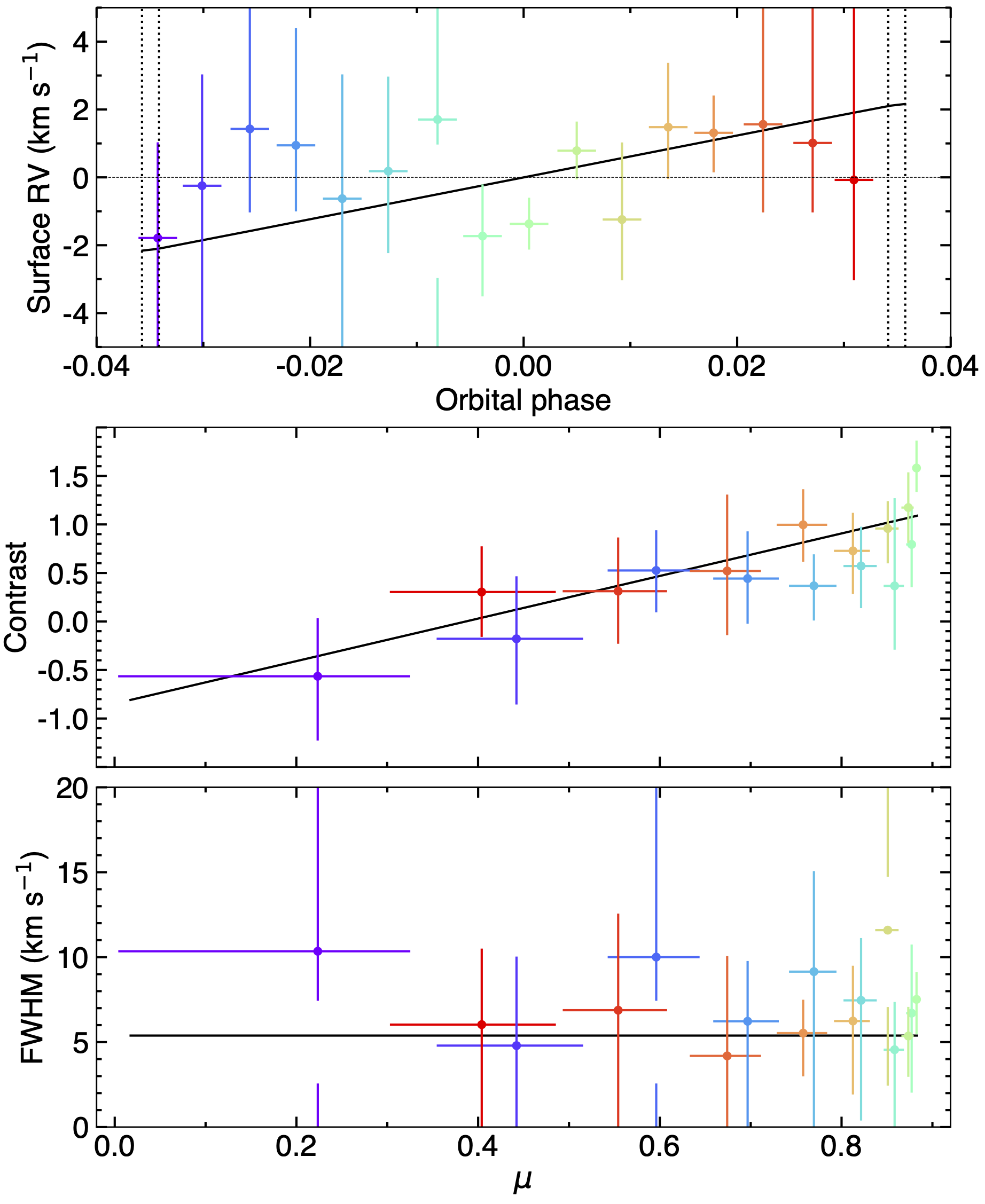}
\centering
\caption[]{Properties of the stellar surface regions occulted by HD\,3167 b (same details as in Fig.~\ref{fig:PropLoc_HD3167c}). The local stellar line cannot be detected in individual exposures, as can be seen from the broad HDI of the line model parameters. Nonetheless, the RV HDIs are not entirely set by the uniform prior, showing that individual exposures contain information on the local stellar line that can be exploited via a global fit to the transit time series. We thus emphasize that the solid black lines models were not fitted to the properties loosely constrained by each individual exposure, but were derived from the global fit to the CCF$_\mathrm{intr}$ time series. }
\label{fig:PropLoc_HD3167b}
\end{figure}

Despite the use of VLT/ESPRESSO to observe the transit of HD\,3167b, the planet is so small that the classical RRM approach fails at detecting and constraining the local stellar line in individual exposures. This can be understood when looking at the results of the MCMC exploration (Fig.~\ref{fig:PDF_HD3167b_indiv}). Most exposures show ill-defined PDFs, or broad with respect to the prior ranges, especially for the local RVs. Even the few exposures near the center of the transit, which have the largest S/N on the night and show better-defined PDFs, still poorly constrain the local RVs (Fig.~\ref{fig:PropLoc_HD3167b}). Nonetheless, information on the local stellar line can still be present even when it is hidden by noise. Binning the consecutive exposures could unveil the local line by increasing the S/N, but at the cost of blurring the signal. The motivation behind the RMR approach, as detailed in the next section, is to circumvent our inability to measure the stellar line properties in individual exposures by fitting all the exposures together, without modifications.

The MCMC fit to individual exposures, along with the analysis of the associated PDFs, remains a useful step in assessing the quality of each CCF$_\mathrm{intr}$ and determining which model best fits the local line profile variations. Here, the last exposure does an especially poor job in constraining the model parameters (Fig.~\ref{fig:PDF_HD3167b_indiv}), likely because it is only partly in-transit, and thus it was excluded from further analysis.

\subsection{Step 3: Global transit analysis}
\label{sec:step3}

The main novelty of the RMR approach lies in fitting all CCF$_\mathrm{intr}$ together with a joint model, rather than fitting them individually. The stellar line profile can be described by a variety of analytical models (e.g., Gaussian; double Gaussian, \citealt{Bourrier_2018_Nat,Bourrier2020_HEARTSIII}). The properties describing the line profile can be kept constant in all exposures or be set by parametric models as a function of a given coordinate parameter. The RV centroids of the theoretical lines are set by the surface RV model described in \citet{Cegla2016} and \citet{Bourrier2017_WASP8}. It accounts for the projected stellar rotational velocity field (plus convective blueshift and differential rotation when relevant) and its blur over the region occulted by the planet for a given exposure. The time series of theoretical stellar lines is convolved with the instrumental response and then fitted to all CCF$_\mathrm{intr}$ together, using \textit{emcee} MCMC to sample the PDFs of the model parameters. Derived values and uncertainties for the model parameters are again defined using medians and HDIs. We note that modeling the stellar line before convolution allows for a more direct comparison between different instruments.

The main interest of the RMR technique is that it boosts the S/N of the occulted stellar line, at first order by the number of in-transit exposures, thus increasing the possibility of constraining the planet path even when the line is not detectable in individual exposures. The RMR approach further exploits the full extent of information contained in the data, as it fits the stellar line profiles rather than just their centroids (as in the traditional RRM approach). Below, we first show how we validated the RMR approach by comparing its results on HD\,3167\,c with those from the traditional RRM approach, and we then demonstrate its importance for small planets via its application to HD\,3167\,b. 

We used a Gaussian profile to model the local stellar lines occulted by both planets. Given that the analysis of individual exposures in Step 2 (Sect.~\ref{sec:step2}) hinted at possible changes in the line shapes along the transit chords, we explored polynomial variations of the contrast or FWHM with $\mu$ = $\cos\theta$ (where $\theta$ is the center-to-limb angle) and used the Bayesian information criterion (BIC, \citealt{Schwarz1978}, \citealt{Liddle2007}) to determine the best polynomial order (see \citealt{Kass1995} for the details of the BIC comparison). The data are not precise enough to search for second-order RV effects caused by the stellar surface motion, and we thus assumed solid-body rotation for the star. Jump parameters for the MCMC are the polynomial coefficients for the contrast and FWHM, the sky-projected obliquity $\lambda$ and stellar rotational velocity, $v$\,sin\,$i_*$. Uniform priors are set on the model parameters, which are uninformative for the contrast and FWHM coefficients (i.e., over a much larger range than their range of variations), between [0 - 10]\,km\,s$^{-1}$ for $v$\,sin\,$i_*$, and over its definition range ([-180 , 180]$^{\circ}$) for $\lambda$. The model was fitted to the CCF$_\mathrm{intr}$ over the range of [-50 , 50]\,km\,s$^{-1}$ in the star rest frame.

\subsubsection{Global analysis for HD\,3167c}

\begin{figure*}
\begin{minipage}[h!]{\textwidth}
\includegraphics[trim=0cm 0cm 0cm 0cm,clip=true,width=0.8\columnwidth]{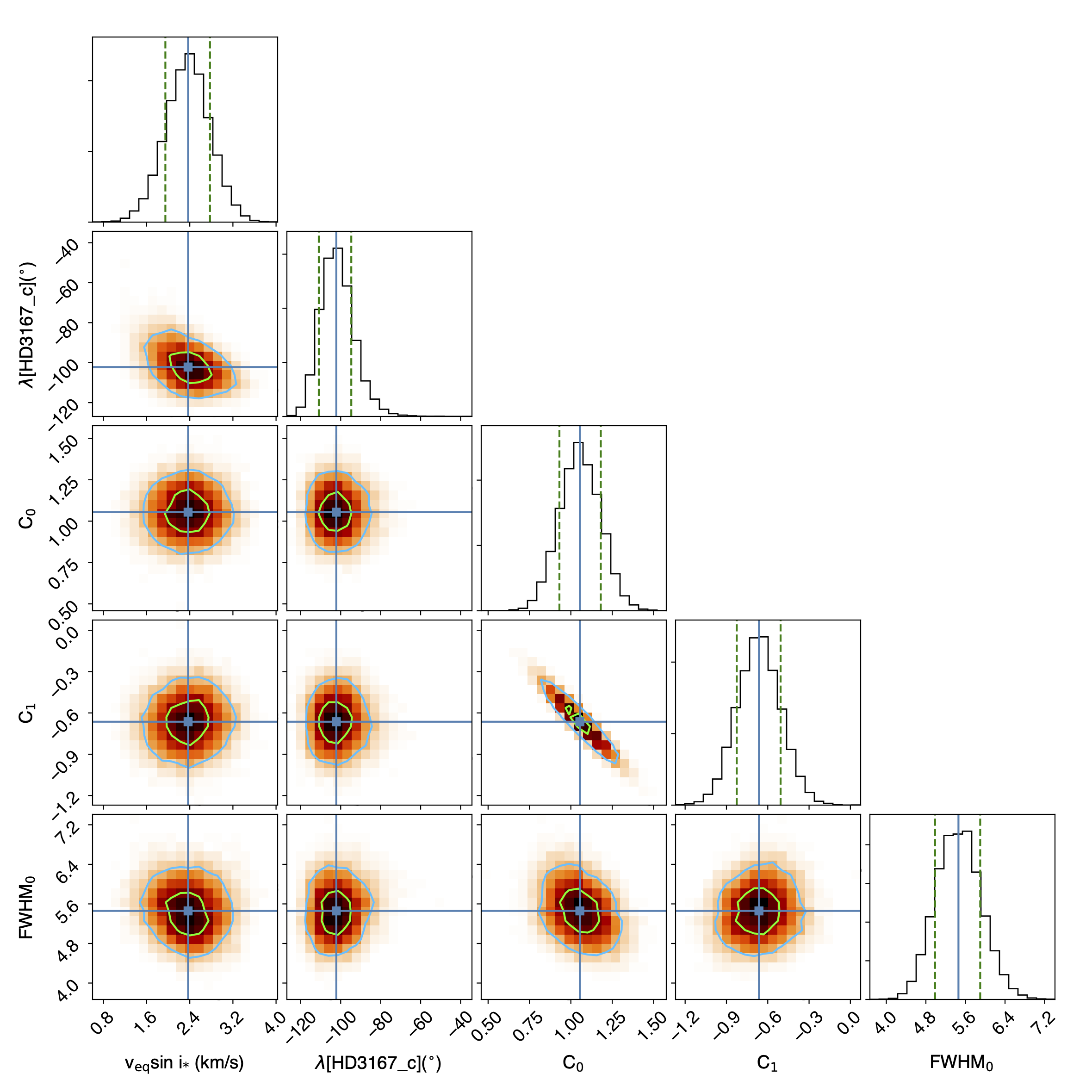}
\centering
\end{minipage}
\caption[]{Correlation diagrams for the PDFs of the RMR model parameters for the HD\,3167c transit. C$_\mathrm{i}$ indicate polynomial coefficients for the contrast model (see text). Green and blue lines show the 1 and 2$\sigma$ simultaneous 2D confidence regions that contain, respectively, 39.3\% and 86.5\% of the accepted steps. 1D histograms correspond to the distributions projected on the space of each line parameter, with the green dashed lines limiting the 68.3\% HDIs. The blue lines and squares show the median values.}
\label{fig:PDF_HD3167c}
\end{figure*}

\begin{figure}
\includegraphics[trim=0cm 0cm 0cm 0cm,clip=true,width=\columnwidth]{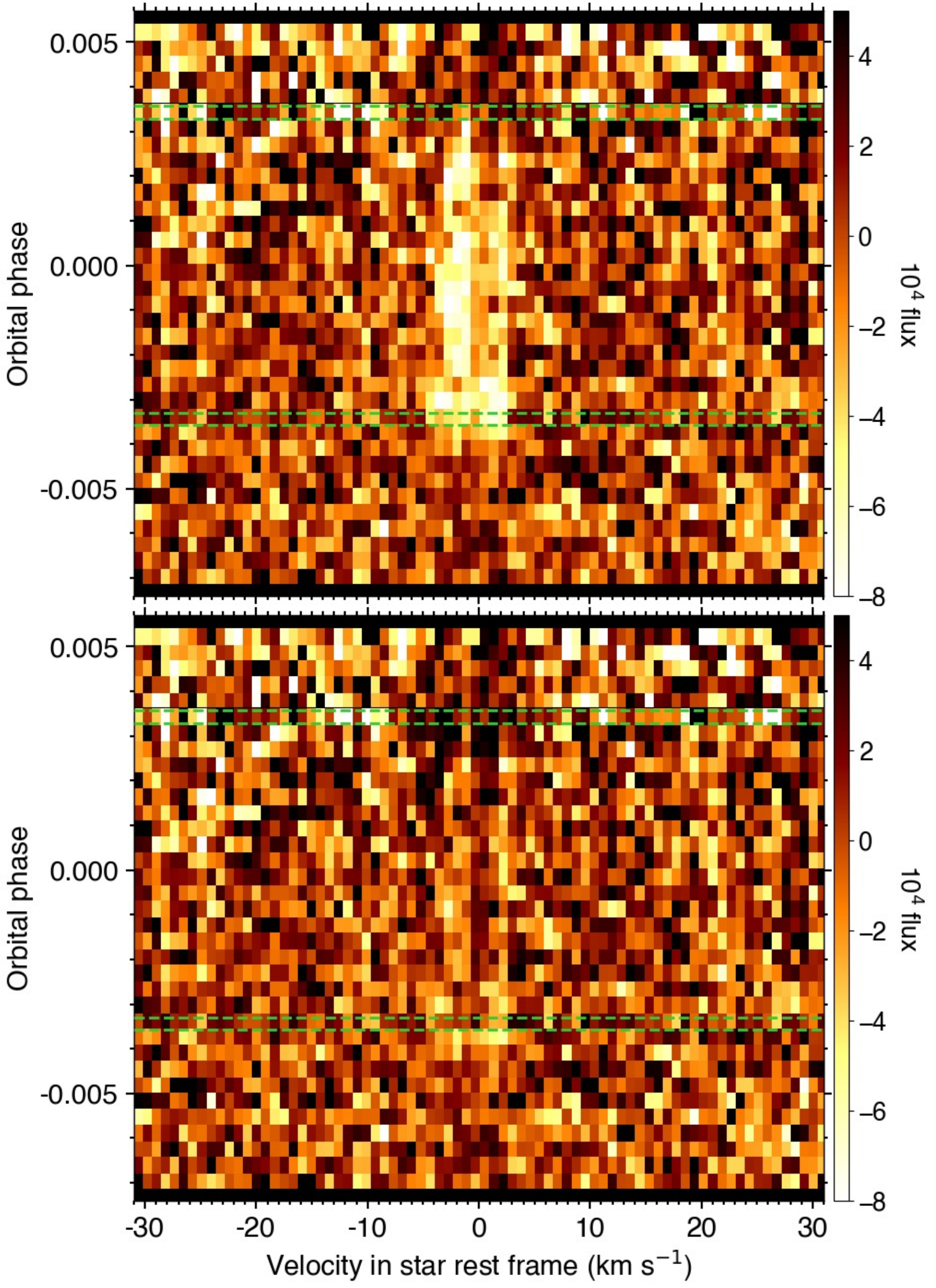}
\centering
\caption[]{Maps of the out-of-transit residual CCF$_\mathrm{loc}$, and of the in-transit residuals between CCF$_\mathrm{intr}$ and their best-fit model (from the joint fit to both transit datasets), during the HD\,3167c visit. Transit contacts are shown as green dashed lines. In the top panel, only the continuum of the CCF$_\mathrm{intr}$ model was subtracted to highlight the occulted stellar lines. In the bottom panel, the full line model was subtracted.}
\label{fig:Res_map_HD3167c}
\end{figure}

The PDFs for the best-fit parameters are shown in Fig.~\ref{fig:PDF_HD3167c}. They are well-defined with symmetrical profiles, so that HDIs are here equivalent to quantile-based credible intervals. Stellar lines along the transit chord of HD\,3167c are best reproduced with a linear variation of the contrast with $\mu$ and a constant FWHM. This model improves upon the BIC by ten points compared to a constant contrast, and by six points compared to a linear variation of the FWHM with $\mu$. Contrast and FWHM of the best-fit line model convolved with the HARPS-N response are shown in Fig.~\ref{fig:PropLoc_HD3167c}. They are in agreement with the properties derived from individual CCF$_\mathrm{intr}$, showing that the RMR approach captures  the line variations along the transit chord well. This is also visible in the residual map between the CCF$_\mathrm{intr}$ and their best-fit models (Fig.~\ref{fig:Res_map_HD3167c}). The best-fit FWHM is 5.46$\pm$0.45\,km\,s$^{-1}$, corresponding to 6.05$\pm$0.41\,km\,s$^{-1}$ after accounting for instrumental convolution. The model contrast increases from the center of the transit chord toward the limbs. Given the highly misaligned orbit of HD\,3167c this could imply a deepening of the lines with increasing latitude, \\textit{i.e.} from stellar equator to stellar poles (e.g., due to the stellar magnetic field, see a similar case in \citealt{Bourrier2017_WASP8}). The exact dependence of this contrast variation with stellar latitude is, however, linked to the unkown inclination of the stellar spin axis with respect to the observer.

We find that $\lambda$ and $v$\,sin\,$i_*$ are not correlated with the contrast and FWHM of the fitted lines, so that their best-fit values do not depend on the line model. We derive $\lambda$ = -102.2$\stackrel{+7.5}{_{-8.7}}^{\circ}$ and $v$\,sin\,$i_*$ = 2.37$\pm$0.42\,km\,s$^{-1}$. These results are consistent within 1$\sigma$ with the values derived by \citet{Dalal2019}, using three different techniques, including the traditional RRM approach, thus validating the RMR approach. In RM analysis $\lambda$ and $v$\,sin\,$i_*$ can be degenerate with the scaled semi-major axis $a/R_{*}$ and orbital inclination $i_{p}$. We allowed these parameters to vary (see \citealt{Bourrier2020_HEARTSIII}), using values from C17 as priors. The resulting fit, however, increases the BIC by 15 points and yields PDFs for $a/R_{*}$ and $i_{p}$ that match their priors, showing that these parameters cannot be better constrained by the present data set.

\subsubsection{Global analysis for HD\,3167b: Detection}
\label{sec:step3_b}

\begin{figure*}
\begin{minipage}[h!]{\textwidth}
\includegraphics[trim=0cm 0cm 0cm 0cm,clip=true,width=0.8\columnwidth]{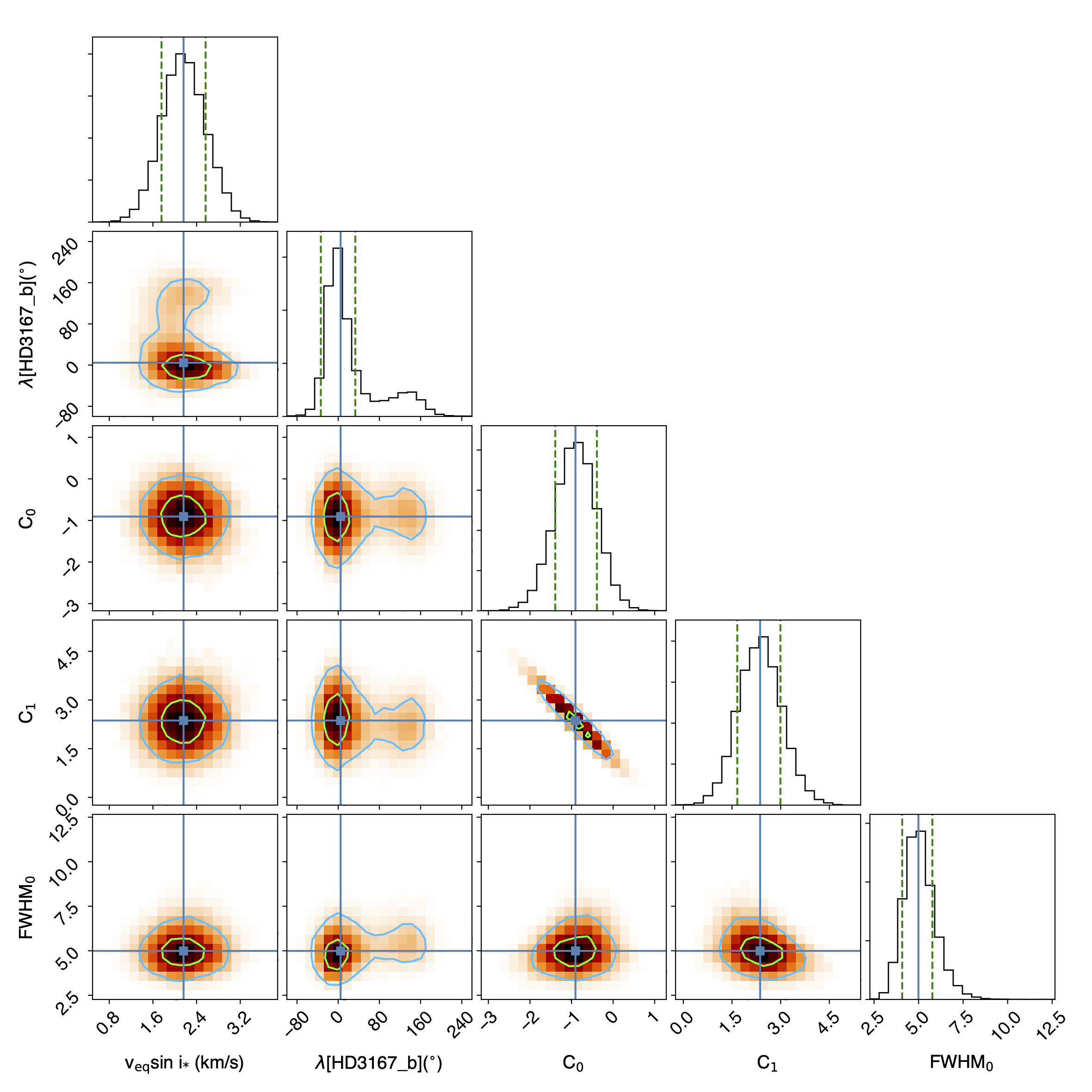}
\centering
\end{minipage}
\caption[]{Correlation diagrams for the PDFs of the RMR model parameters for the HD3167b transit (same details as in Fig.~\ref{fig:PDF_HD3167c}).}\label{fig:PDF_HD3167b}
\end{figure*}

\begin{figure}
\includegraphics[trim=0cm 0cm 0cm 0cm,clip=true,width=\columnwidth]{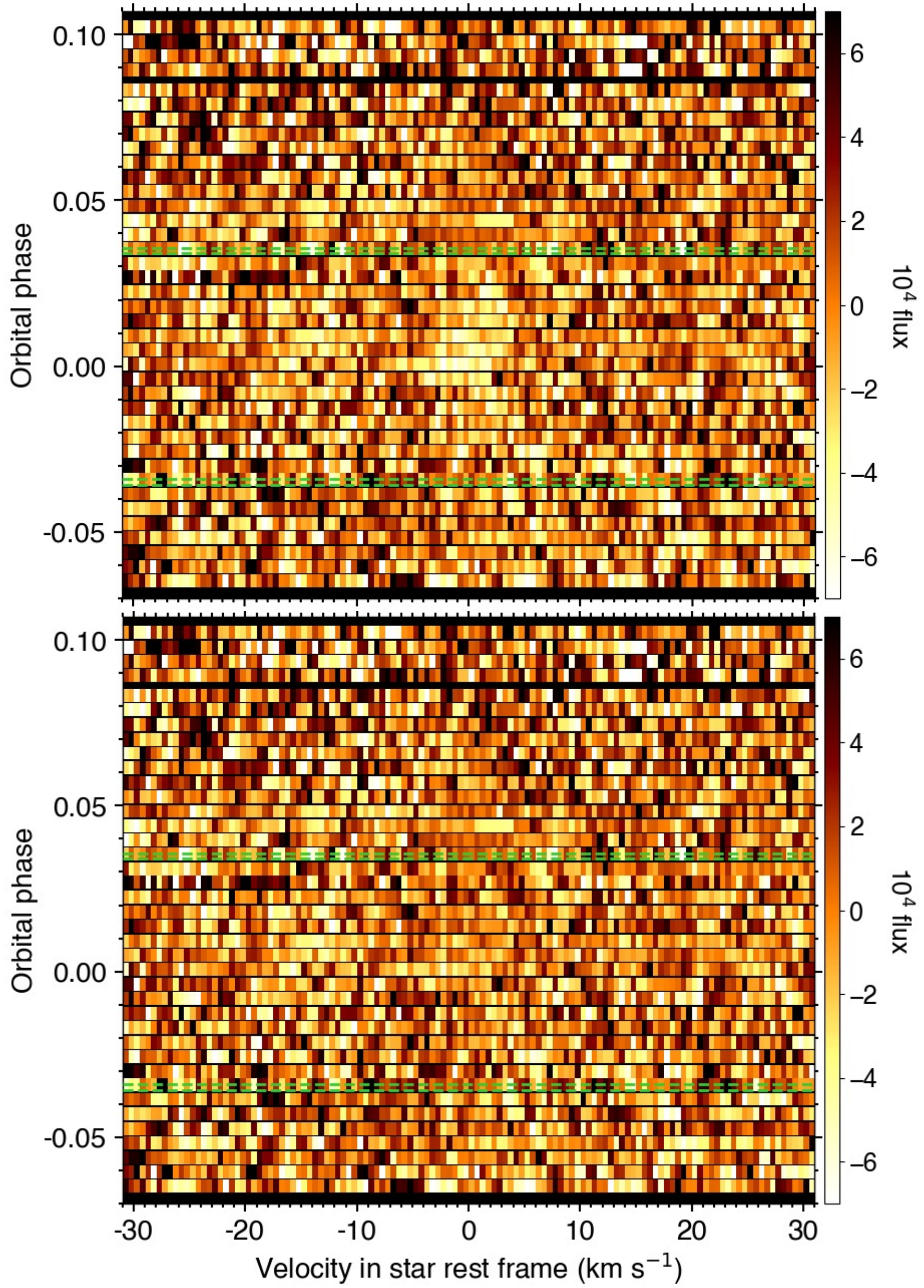}
\centering
\caption[]{Residual maps during the HD\,3167b visit (same details as in Fig.~\ref{fig:Res_map_HD3167c}).}
\label{fig:Res_map_HD3167b}
\end{figure}

We exploited results from the HD\,3167c fit to refine the prior on $v$\,sin\,$i_*$, using a normal distribution with center 2.37\,km\,s$^{-1}$ and width 0.42\,km\,s$^{-1}$. Residuals between the CCF$_\mathrm{intr}$ and the best-fit line models are shown in Fig.~\ref{fig:Res_map_HD3167b}. The corresponding PDFs for the model parameters are shown in Fig.~\ref{fig:PDF_HD3167b}. Apart from $\lambda,$ these are well defined with mostly symmetrical profiles. Stellar lines along the transit chord are best reproduced with a linear variation in contrast with $\mu$ and a constant FWHM (Fig.~\ref{fig:PropLoc_HD3167b}), improving the BIC by six points compared to a constant contrast or to a linear variation in FWHM with $\mu$. The best-fit FWHM is 5.00$\stackrel{+0.78}{_{-0.91}}$\,km\,s$^{-1}$ (corresponding to 5.38$\stackrel{+0.72}{_{-0.85}}$\,km\,s$^{-1}$ for the line observed with ESPRESSO), in good agreement with the value derived from the HARPS-N transit even though the ESPRESSO and HARPS-N CCF masks are not the same. In contrast to HD\,3167c, the derived line contrast decreases from the center of HD\,3167b transit chord toward the limbs. This could indicate that the contrast variations observed for the two planets do not depend primarily on $\mu$, but on stellar latitude instead. 

Indeed, one of the main results from our study is that HD\,3167b and HD\,3167c transit along nearly perpendicular tracks across the stellar surface, and thus occult very different stellar latitudes. The RMR approach allows us to detect the RM signal of HD\,3167\,b, as shown by the PDF of the model obliquity, with $\lambda$ = 5$\stackrel{+28}{_{-38}}^{\circ}$ (Fig.~\ref{fig:PDF_HD3167b}). This is the first clear detection of a spectroscopic RM signal for a planet smaller than 2$\,R_{\Earth}$ (the second smallest being Pi Men c, 2.1$\,R_{\Earth}$, \citealt{Kunovac2021}), showing the potential of the RMR approach to unveil the orbital architectures of the smallest exoplanets. As with HD\,3167c, we note that $\lambda$ is not correlated with the line shape properties (Fig.~\ref{fig:PDF_HD3167b}) and, accordingly, its value does not depend on the contrast and FWHM models.

The ESPRESSO data do not constrain much the stellar rotational velocity compared to the prior from the HARPS-N fit, with $v$\,sin\,$i_*$ = 2.16$\stackrel{+0.41}{_{-0.39}}$\,km\,s$^{-1}$. They also do not constrain the scaled semi-major axis $a/R_{*}$ and orbital inclination $i_{p}$ of HD\,3167b better than the priors from C17, as letting these parameters free to vary increased the BIC by 16 points with no change in their distributions.

\subsubsection{Global analysis for HD\,3167b: Validation}

We carried out several tests to assess the validity of the HD\,3167b RM detection.

First, we tested the null hypothesis, namely, the absence of a detectable stellar line from the regions occulted by HD\,3167b. The corresponding model, a constant flux in all CCF$_\mathrm{intr}$, yields a BIC larger than the best-fit model by 80 points, unambiguously displaying the presence of a signal in the data.  
   
Then, to ensure that the detected signal is not a spurious feature arising from noise, we performed the same fits as in Sect.~\ref{sec:step3_b}, but in the blue and in the red continuum of the CCF$_\mathrm{intr}$. The nominal fit was carried out over the range of [-50 ; 50]\,km\,s$^{-1}$ in the star rest frame. For the ``blue'' test, we artificially set  the star rest frame at -160\,km\,s$^{-1}$, that is, we shifted all CCF$_\mathrm{intr}$ by 160\,km\,s$^{-1}$ to perform the fit over the original range [-210 ; -110]\,km\,s$^{-1}$. For the ``red'' test, all CCF$_\mathrm{intr}$ were shifted by -160\,km\,s$^{-1}$ , so that the fit was performed over [110 ; 210]\,km\,s$^{-1}$. We show in Appendix \ref{apn:corr_cont}, the correlation diagrams for these two fits. They converge toward model stellar lines with contrast and FWHM consistent with null values, and a roughly uniform PDF for $v$\,sin\,$i_*$ and $\lambda$. This is the expected result when fitting a region of constant flux with photon noise and no spurious features, giving us confidence that there are no systematics in the CCF$_\mathrm{intr}$ that could mimic the detected signal. 

Finally, to further ensure that the detected signal is not a spurious feature we performed a fit on the residual CCF$_\mathrm{loc}$ time series obtained after the transit, which covers a phase range of same duration as the fitted transit window. As can be seen in Fig.~\ref{fig:Corr_diag_postTR}, the fit again converges toward a model stellar line with null contrast and FWHM, as expected from a constant flux with white noise.


\subsection{Multi-planet fit}
\label{sec:multifit}

\begin{figure*}
\begin{minipage}[h!]{\textwidth}
\includegraphics[trim=0cm 0cm 0cm 0cm,clip=true,width=\columnwidth]{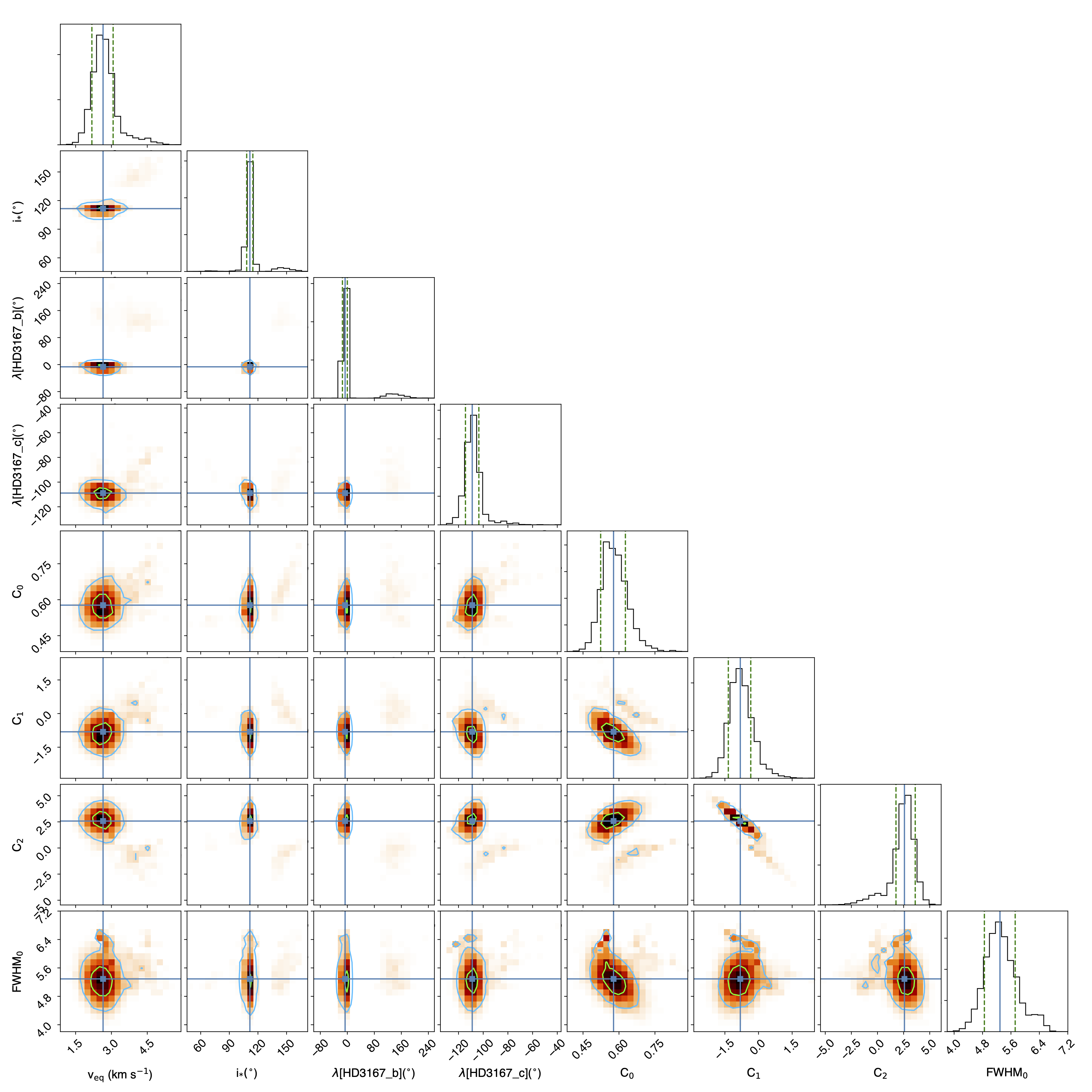}
\centering
\end{minipage}
\caption[]{Correlation diagrams for the PDFs of the RMR model parameters from the joint fit to the HD\,3167b and HD\,3167c transits (same details as in Fig.~\ref{fig:PDF_HD3167c}).}
\label{fig:PDF_HD3167_joint}
\end{figure*}

While we cannot exclude this possibility, we found no evidence for an instrumental origin for the contrast variations measured for HD\,3167b and HD\,3167c. The transit light curves used to scale the CCF$_\mathrm{DI}$ (Sect.~\ref{sec:step1}) have an impact on the derived local contrasts, but varying the transit depth within its uncertainties does not suppress the contrast variations. The different trends in local line contrast for the two planets can, however, be reconciled within the frame of an intriguing scenario. Due to the different obliquities of their orbits, HD\,3167b and HD\,3167c  effectively occult regions of very different latitudes across the stellar surface. While no meridional variations in line contrast are observed over the solar surface, HD\,3167 is a K-type star with a super-Earth orbiting within its corona, at only four times the stellar radius. The measured line contrast variations can be explained by a common trend if we assume that the stellar surface properties change with latitude, either due to the star itself or because of interactions with its close-in companion (e.g., \citealt{Bourrier2018_FUV}). In the framework of this scenario, we applied a joint RMR fit to the two transit datasets, using a common intrinsic line model. The line positions are set by the theoretical transit chords of each planet, controlled by their respective sky-projected obliquities, and by the theoretical surface velocity field of the star, now controlled independently by the stellar equatorial velocity, $v_\mathrm{eq}$, and the stellar inclination, $i_{\star}$. We fit a common FWHM for the intrinsic Gaussian line in the ESPRESSO and HARPS-N data, since the independent fits to each dataset yielded widths consistent within their uncertainties (Sect.~\ref{sec:step3}). Inspired by differential rotation models, we assume that the intrinsic line contrast varies as a polynomial function of $sin(\theta_{lat})^2$, where $\theta_{lat}$ is the latitude relative to the stellar equator (we also assume that variations of the contrast with $\mu$ are negligible compared to its latitudinal variations). Since this model depends on absolute latitude, the stellar inclination is fitted within [0 , 180]$^{\circ}$ (where $i_{\star}$ is the angle between the line-of-sight and stellar spin axis, see \citealt{Cegla2016} for details on the coordinate system). We set uniform priors between [0 - 10]\,km\,s$^{-1}$ for $v_\mathrm{eq}$, and between [-1,1] for $cos(i_{\star})$ (used as jump parameter for the stellar inclination). The PDFs for the model parameters, sampled as in Sect.~\ref{sec:step3}, are shown in Fig.~\ref{fig:PDF_HD3167_joint}.

\begin{figure}
\includegraphics[trim=1.5cm 0cm 1.2cm 5cm,clip=true,width=\columnwidth]{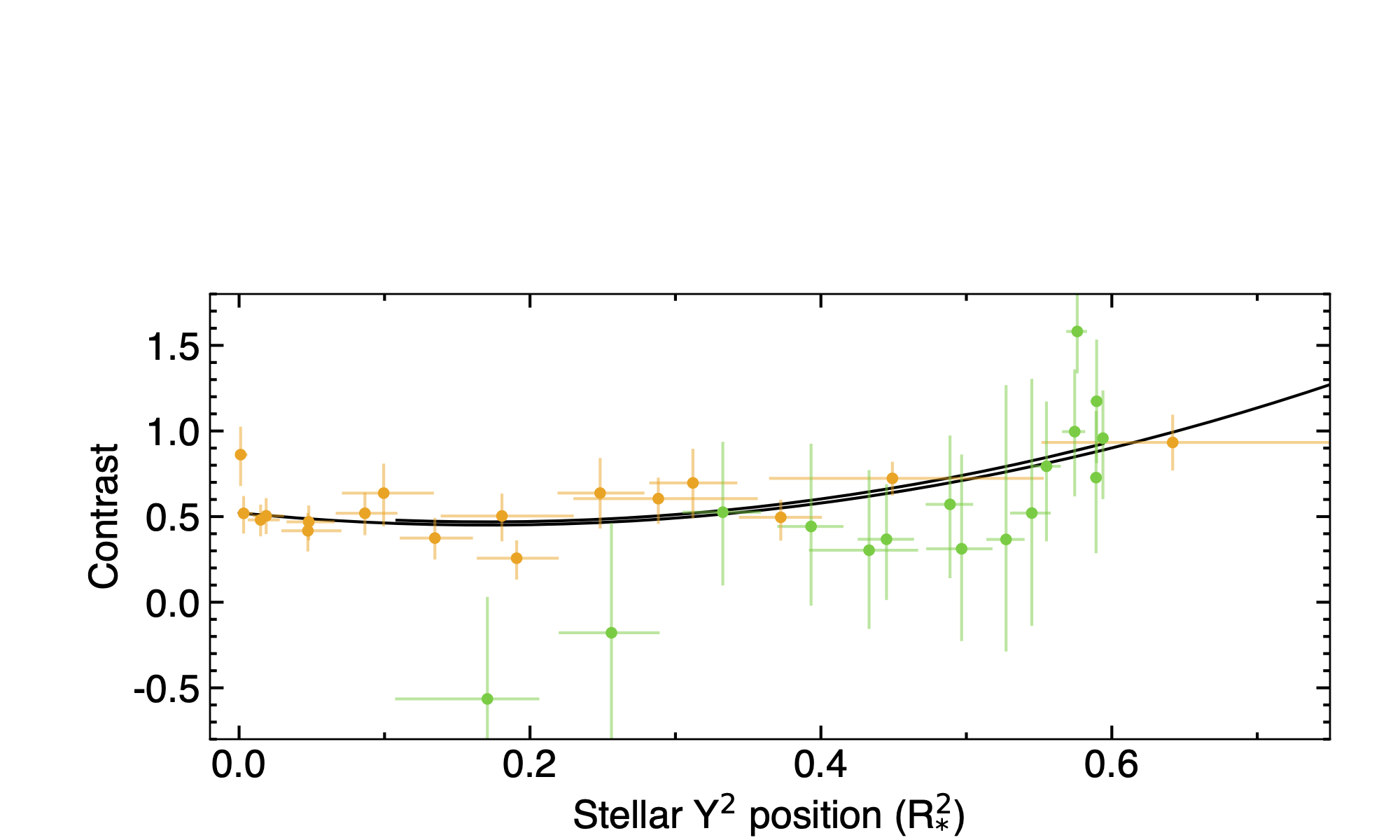}
\centering
\caption[]{Local contrast of the stellar lines occulted by HD\,3167b (green) and HD\,3167c (orange), derived from the fits to individual exposures (Sect.~\ref{sec:step2}, Figs.~\ref{fig:PropLoc_HD3167c}, and \ref{fig:PropLoc_HD3167b}). Solid black lines show the best-fit model for the contrast variations as a function of the squared latitude coordinate $sin(\theta_{lat})^2$ (ranging from 0 at the stellar equator to 1 at the poles), after convolution by the HARPS-N and ESPRESSO LSFs. We note that the models were derived from the RMR fit to the combined datasets. }
\label{fig:ctrst_ystar2}
\end{figure}

We find that the different line contrast variations measured for both planets can be explained by a common quadratic function of $sin(\theta_{lat})^2$ (Fig.~\ref{fig:ctrst_ystar2}), providing a better BIC than a linear function by eight points, and than a third-order polynomial by five points. In this scenario, the stellar spin has to be slightly inclined ($i_{\star}$ = 111.6$\stackrel{+3.1}{_{-3.3}}^{\circ}$), so that HD\,3167b transits closer to the stellar pole (Fig.~\ref{fig:System_view}). Breaking the degeneracy on $v$\,sin\,$i_*$ allows us to derive the stellar equatorial rotational velocity $v_\mathrm{eq}$ = 2.65$\stackrel{+0.47}{_{-0.42}}$\,km\,s$^{-1}$. Accounting for the uncertainties on R$_\star$ = 0.850$\pm$0.020\,R$_\odot$ (derived from a spectroscopic analysis of the ESPRESSO data, following the approach from \citealt{daSilva2006}), this result further yields the equatorial rotation period of the star as $P_\mathrm{eq}$ = 16.2$\stackrel{+2.8}{_{-2.7}}$\,d.

\begin{figure}
\includegraphics[trim=0cm 0cm 0cm 0cm,clip=true,width=\columnwidth]{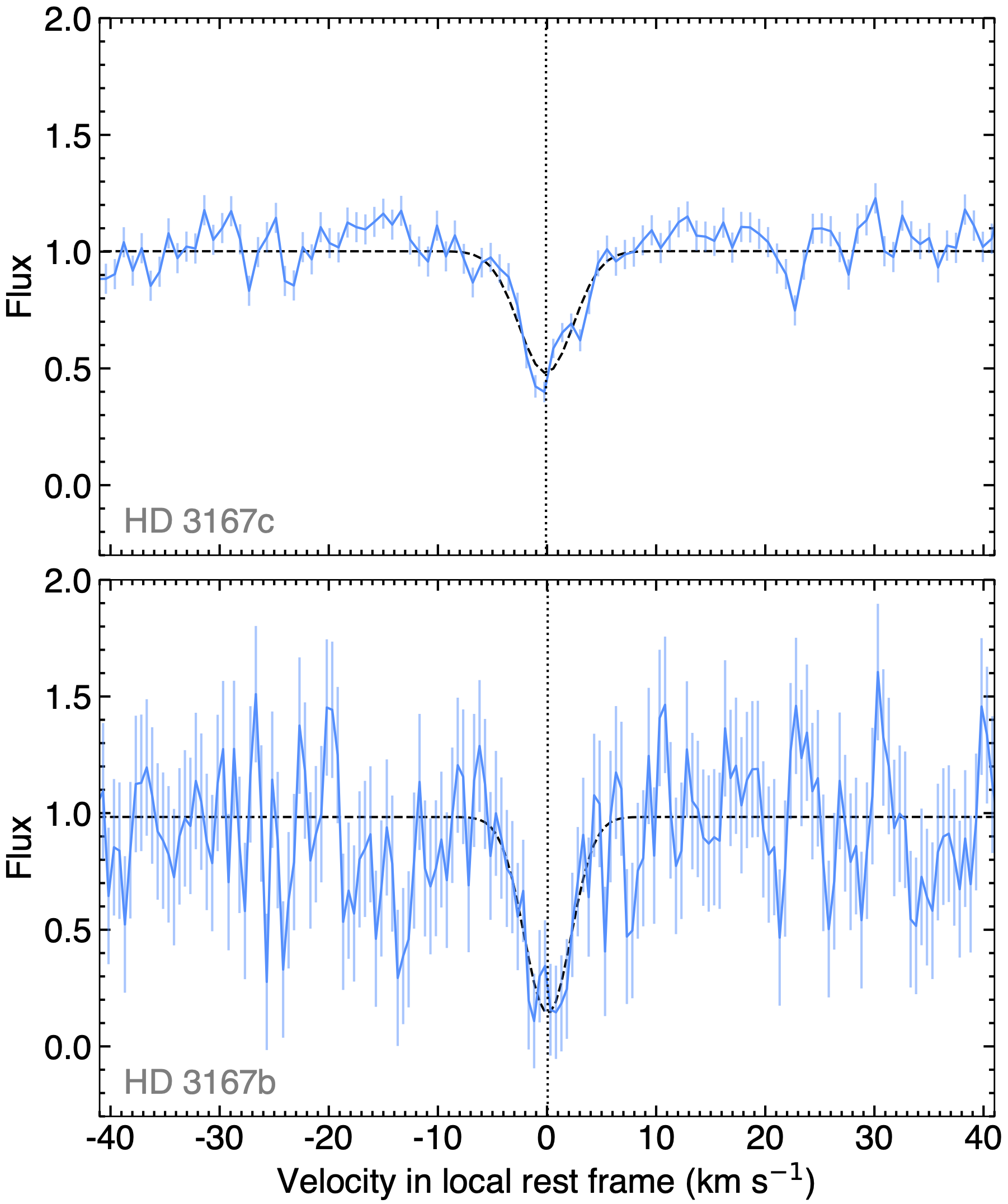}
\centering
\caption[]{Average local stellar lines along the transit chords of HD\,3167c (top panel) and HD\,3167b (bottom panel). The same flux scale is used to highlight the difference in line contrast, arising from the different stellar regions probed by the planets, and the difference in signal quality, due to the different planet sizes. Dashed black lines show best-fit Gaussian profiles to the stellar lines.}
\label{fig:Local_Masters}
\end{figure}

As with the fits to each dataset, the obliquities are not correlated with the line contrast, giving us confidence in the derived $\lambda_\mathrm{b}$ = -6.6$\stackrel{+6.6}{_{-7.9}}^{\circ}$ and $\lambda_\mathrm{c}$ = -108.9$\stackrel{+5.4}{_{-5.5}}^{\circ}$. These values are consistent with those derived from each dataset, but their precision was improved by fitting the two datasets together. We note that local line contrasts from some of the regions occulted by HD\,3167b are outside [0 ; 1], although the largest deviation remains within 2$\sigma$ from this range (Fig.~\ref{fig:ctrst_ystar2}). This could be due to uncertainties in the model light curve used to scale in-transit CCF$_\mathrm{DI}$, as a transit depth larger than its actual value would increase  the local line contrasts artificially. We checked, however, that varying the transit depth of both planets within their 1$\sigma$ uncertainties has no impact on the derived obliquities. We show in Fig.~\ref{fig:Local_Masters} the average of the CCF$_\mathrm{intr}$ measured along each transit chord, after they were aligned using the best-fit surface RV models. Gaussian fits to these average lines yield contrasts of about 52\% for HD\,3167c and 86\% for HD\,3167b, illustrating the fact that the two planets occult stellar regions with different properties.

\section{Orbital architecture of the HD\,3167 system in context}
\label{sec:orb_arch}

\begin{figure}
\includegraphics[trim=0cm 0cm 0cm 0cm,clip=true,width=\columnwidth]{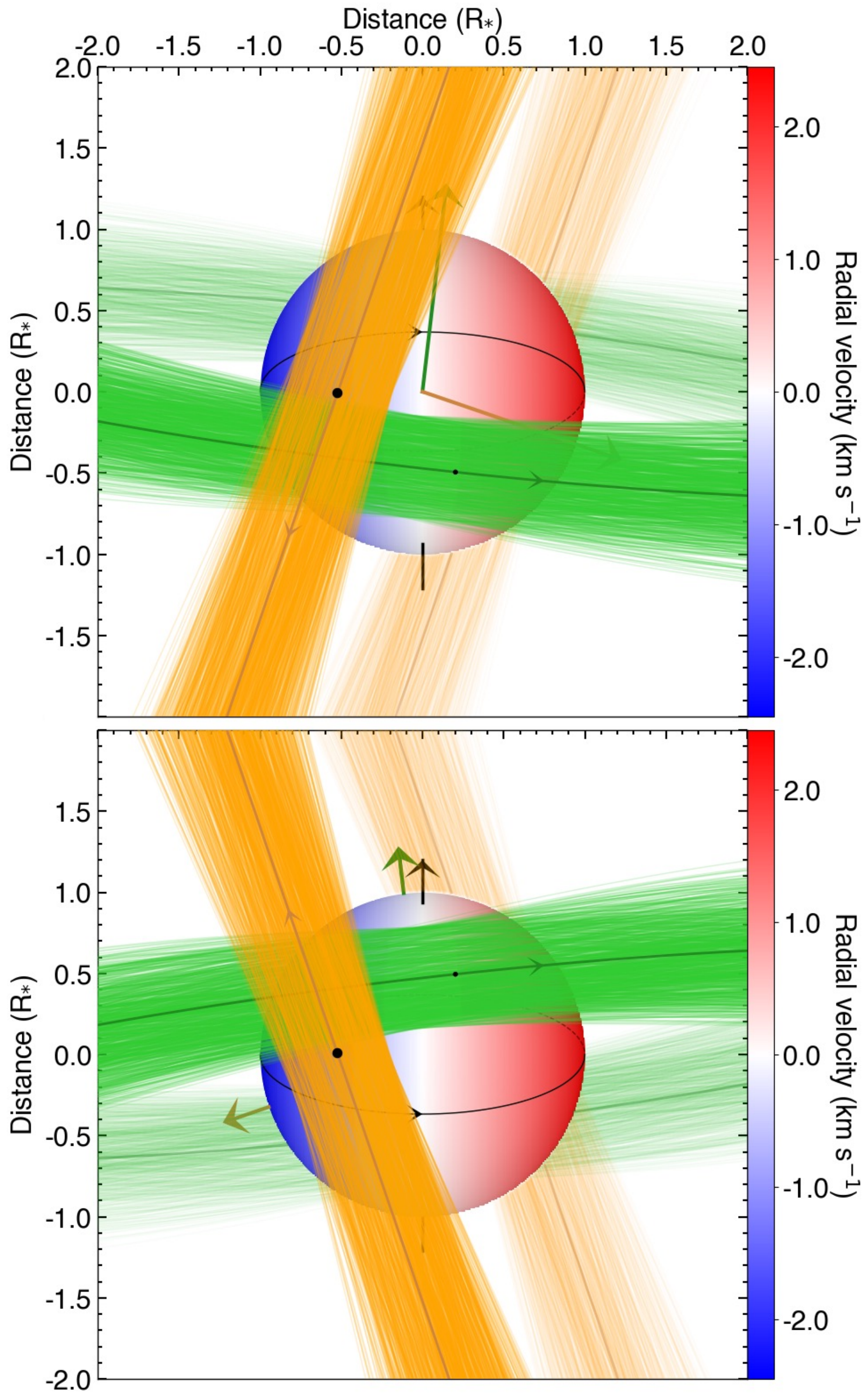}
\centering
\caption[]{Projection of HD\,3167 in the plane of sky for the best-fit orbital architecture (the top and bottom panels show configurations A and B, respectively; see text). The stellar spin axis is displayed as a black arrow extending from the north pole. The stellar equator is represented as a black line. The stellar disk is colored as a function of its surface RV field. Normals to the orbital planes of HD\,3167b and HD\,3167c are shown as green and brown arrows, respectively. Thick solid curves with corresponding colors represent the best-fit orbital trajectories. The thin lines surrounding them show orbits obtained for orbital inclination, semi-major axis and sky-projected obliquity values drawn randomly within 1$\sigma$ from their probability distributions. The star, the planets HD\,3167b and HD\,3167c (black disks), and their orbits are shown to scale. HD\,3167d, which orbits in between HD\,3167b and HD\,3167c, is not shown because of its unknown inclination.}
\label{fig:System_view}
\end{figure}

The main architectural properties we derived on the HD\,3167 system are reported in Table \ref{tab:archi_derived}. The observables in our study, namely, the RV position and contrast of the local stellar lines occulted by the planets, leave a degeneracy between two possible orbital configurations shown in Fig.~\ref{fig:System_view}: A ($i_{b}$, $\lambda_{b}$, $i_{c}$, $\lambda_{c}$, $i_{\star}$) and B ($\pi$-$i_{b}$, -$\lambda_{b}$, $\pi$-$i_{c}$, -$\lambda_{c}$, $\pi$-$i_{\star}$). These two configurations, however, yield the same values for the 3D spin-orbit angles of the planets $\psi$ = arccos(sin\,$i_{\star}$ cos\,$\lambda$ sin\,$i_\mathrm{p}$ + cos\,$i_{\star}$ cos\,$i_\mathrm{p}$), and for the mutual inclination between their orbital planes $i_\mathrm{mut}$ = arccos(cos\,$i_{b}$ cos\,$i_{c}$ + cos\,$\Omega$ sin\,$i_{b}$ sin\,$i_{c}$). Here, $\Omega$ is the difference between the longitudes of the planets' ascending nodes, corresponding to $\lambda_{c}$-$\lambda_{b}$ (A) or -$\lambda_{c}$+$\lambda_{b}$ (B).

\begin{figure}
\includegraphics[trim=0cm 0cm 0cm 0cm,clip=true,width=\columnwidth]{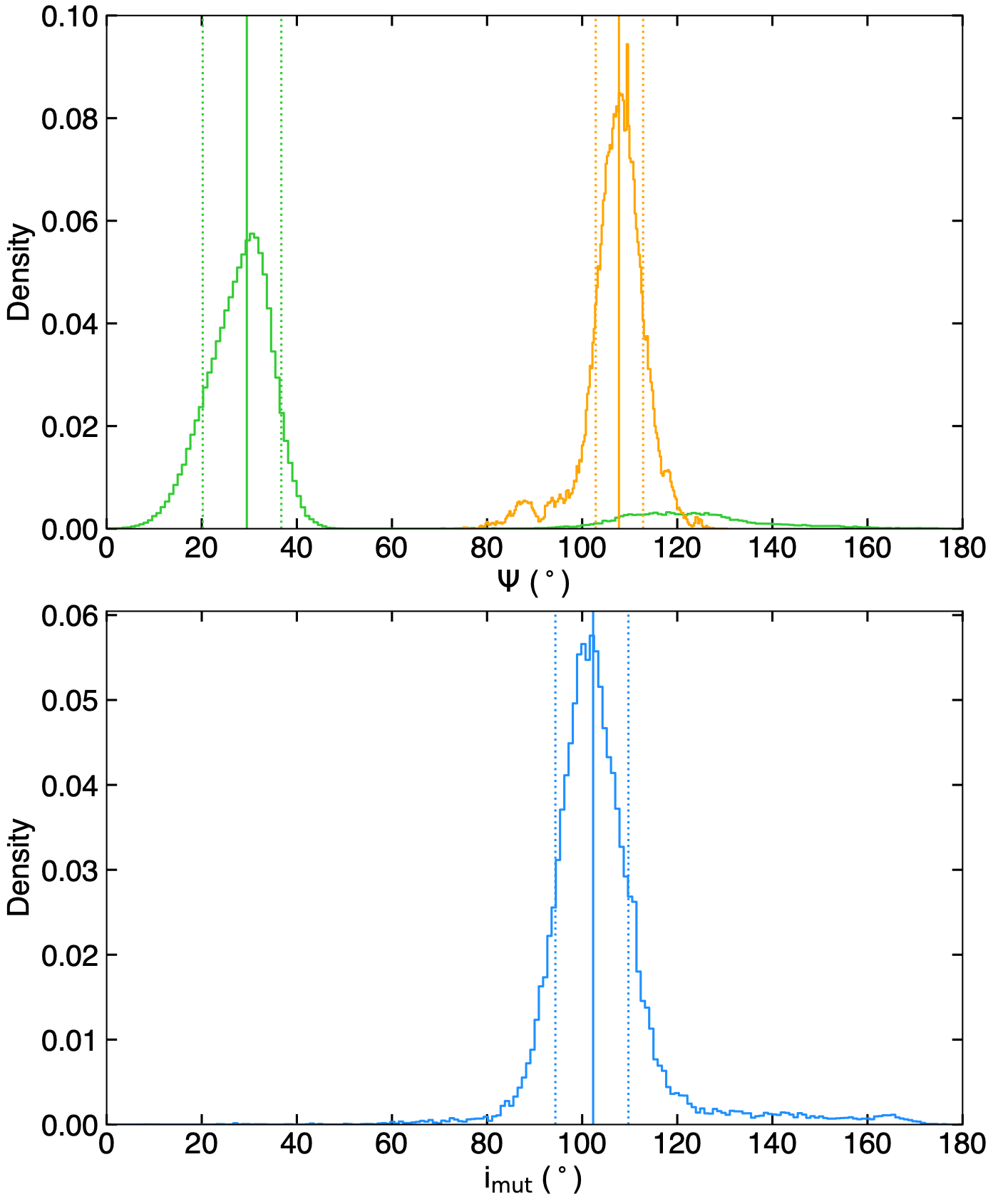}
\centering
\caption[]{PDFs of the 3D spin-orbit angles (\textit{top panel}) for HD\,3167b (green) and HD\,3167c (orange), and of the mutual inclination between the two planets (blue, \textit{bottom panel}). Solid lines indicate the PDF median values, with dashed lines showing the 1$\sigma$ HDIs. }
\label{fig:Psi_imut}
\end{figure}

We combined the probability distributions of $i_\mathrm{p}$, $\lambda$, and $i_{*}$ to calculate the PDFs of $\psi$ and $i_\mathrm{mut}$ (Fig.~\ref{fig:Psi_imut}). We derive $\psi^{\rm b}$ = 29.5$\stackrel{+7.2}{_{-9.4}}^{\circ}$, $\psi^{\rm c}$ = 107.7$\stackrel{+5.1}{_{-4.9}}^{\circ}$, and $i_\mathrm{mut}$ = 102.3$\stackrel{+7.4}{_{-8.0}}^{\circ}$. Thus, HD\,3167b is close to orbiting within the stellar equatorial plane, while HD\,3167c is on a near-polar orbit around the star, and the orbital planes of HD\,3167b and HD\,3167c are nearly perpendicular.\\

Dynamical analysis by C17 and \citet{Dalal2019} showed that HD\,3167c and d are not coplanar but have low mutual inclinations, likely between 2.3 and 21$^{\circ}$. The high spin-orbit angle of HD\,3167c thus implies that both planets are on near-polar orbits around the star. \citet{Dalal2019} further predicted that the stellar spin could not be aligned with the angular momentum of these two planets, which is confirmed by our measurement of $\psi$ for HD\,3167c. These authors concluded that if they formed on aligned orbits, HD\,3167c and HD\,3167d are unlikely to have reached such misaligned orbits by interacting with the star. They proposed instead that the orbital planes of these planets were misaligned via gravitational interactions with an outer inclined companion (\citealt{boue2014b}). This companion should have been massive enough and close enough to HD\,3167c and HD\,3167d that their dynamical evolution was not dominated by gravitational interactions with the star, but not too massive and close that it entered a Kozai resonance with HD\,3167c and destabilized the outer planets. In the intermediate regime, the dynamical evolution of HD\,3167c and HD\,3167d would have been controlled by gravitational interactions with the companion, which gently excited their orbits to high inclinations with respect to the star over secular timescales. 

At such short orbital distance (only four times the stellar radius), the present-day orbit of HD\,3167b is governed by interactions with the star and the dynamical stability of the system does not depend on this planet's inclination (\citealt{Dalal2019}). We propose that HD\,3167b formed with a low enough mass and far enough away from planets c and d that their dynamical evolution was always decoupled, and that HD\,3167b was not influenced by the outer companion. While HD\,3167b is close enough to the star that it might have reached its present orbit due to tidal interactions, the three planets likely initially migrated via disk-driven interactions, remaining close to the stellar equatorial plane. As the star spinned down HD\,3167c and HD\,3167d became more sensitive to the gravitational influence of the outer companion (\citealt{Dalal2019}) and were progressively misaligned, while HD\,3167b remained strongly coupled to the star and kept its primordial alignment. We note that we consider HD\,3167b as aligned following the limit of $\Psi\sim$30\,$^{\circ}$ commonly used in the exoplanet literature (e.g., \citealt{Triaud2010,Albrecht2012}). \citet{Albrecht2021} recently confirmed that exoplanet systems do not span the full range of obliquities, but show a preference for nearly-perpendicular (``misaligned''; $\Psi$ = 80--125\,$^{\circ}$) or low-obliquity (``aligned''; $\Psi\approxinf$ 35\,$^{\circ}$) orbits.\\

\begin{table}[h]
\centering
\caption{Architectural properties of the HD\,3167 system}
\begin{tabular}{lcccc}
\hline
Parameter & Value & Unit\\
\hline
$v$\,sin\,$i_*$ &  2.41$\pm$0.37   &  km\,s$^{-1}$\\
$P_\mathrm{eq}$  &  16.2$\stackrel{+2.8}{_{-2.7}}$   &  days  \\
$\psi^{\rm b}$ & 29.5$\stackrel{+7.2}{_{-9.4}}$  & deg\\
$\psi^{\rm c}$ & 107.7$\stackrel{+5.1}{_{-4.9}}$  & deg\\
$i_\mathrm{mut}$ & 102.3$\stackrel{+7.4}{_{-8.0}}$  & deg\\
\hline
\multicolumn{3}{c}{Configuration A} \\
$i_{\star}$ & 111.6$\stackrel{+3.1}{_{-3.3}}$  & deg\\
$\lambda_\mathrm{b}$ & -6.6$\stackrel{+6.6}{_{-7.9}}$  & deg\\
$\lambda_\mathrm{c}$ & -108.9$\stackrel{+5.4}{_{-5.5}}$  & deg\\
\hline
\multicolumn{3}{c}{Configuration B} \\
$i_{\star}$ & 68.4$\stackrel{+3.3}{_{-3.1}}$  & deg\\
$\lambda_\mathrm{b}$ & 6.6$\stackrel{+7.9}{_{-6.6}}$  & deg\\
$\lambda_\mathrm{c}$ & 108.9$\stackrel{+5.5}{_{-5.4}}$  & deg\\
\hline
\end{tabular}
\label{tab:archi_derived}
\end{table}

It is interesting to compare HD\,3167 with other multi-planet systems with known architectures. HD\,3167 provides an interesting counterpoint to the multi-planet systems Kepler-56 and K2-290\,A. Kepler-56 hosts two inner giant planets on nearly coplanar orbits, but that are misaligned with their star (\citealt{huber2013}). Dynamical simulations by \citet{Gratia2017} show that late scattering with the third more massive outer planet in the system (\citealt{Otor2016}) could have misaligned the inner system with the star. K2-290\,A also hosts a mini-Neptune and a Jupiter-size planet on coplanar orbits (\citealt{Hjorth2021}). Their Rossiter-McLaughlin signatures, analyzed with the classical RV technique, are however consistent with mutually aligned planetary orbits both retrograde with respect to the star. Combined with the presence of a stellar companion on a wide orbit, this finding led \citet{Hjorth2021} to conclude that the most likely scenario for the K2-290\,A system is a primordial misalignment of the protoplanetary disk with the primary star, caused by the gravitational torque from the stellar companion. The dynamical evolution of the HD\,3167's outer planetary system could be similar to what was proposed for Kepler-129 by \citet{Zhang2021}, also in the framework developed by \citet{boue2014b}. The orbits of the two sub-Neptunes around Kepler-129 could have become misaligned as a result of precessing around the orbital normal of a long-period giant planet discovered in the system. The HD\,3167 system -- with outer planets that were likely tilted by secular interactions with an outer companion but an inner planet still nearly aligned with the star -- offers yet another example of a complex dynamical history. These various results illustrate how the analysis of RM signatures in transiting multi-planet systems allows us to distinguish between various primordial or secular misalignment scenarios.

\section{Conclusion}
\label{sec:conclu}

Rossiter-McLaughlin studies of transiting exoplanets yield precious information about their orbital architecture. Recent improvements in analysis techniques have unveiled the signal of smaller exoplanets (e.g., \citealt{Kunovac2021}) around slower rotators (e.g., \citealt{Bourrier_2018_Nat}). Yet even with next-generation spectrographs such as ESPRESSO, the RM signal of super-Earths and smaller planets remain difficult to reach. Building on the latest technique (\citealt{Cegla2016}), we developed the Rossiter-McLaughling effect Revolutions (RMR) approach to exploit even further the information contained in transit time series. This technique relies on fitting together the spectral profiles from all planet-occulted regions with a joint model, with the possibility to combine transit data sets from multiple planets obtained with different instruments. We first validate the RMR approach on HARPS-N observations of the mini-Neptune HD\,3167c, confirming its known RM signal and then apply it to ESPRESSO observations of the super-Earth HD\,3167b, detecting its RM signal for the first time. We emphasize the critical importance of precise ephemeris for RM studies of close-in planets, which requires re-observing their transit regularly (e.g., \citealt{CasasayasBarris2021}). HD\,3167\,b is the smallest exoplanet (1.7\,R$_{\rm Earth}$), with a clear detection of a spectroscopic RM signature (see the cases of 55\,Cnc e, \citealt{bourrier2014b}, \citealt{lopez2014}; and of the TRAPPIST-1 system, \citealt{Hirano2020}). The RMR approach thus opens a new window into the orbital architecture of Earth-size planets, up to now mainly probed via photometric techniques (e.g., \citealt{Kamiaka2019}).

Curiously, the contrast of the stellar lines occulted by HD\,3167b and HD\,3167c display different variations from the center to the limb of the star. While we cannot exclude an instrumental origin or a bias due to the photometric scaling of the data, it is remarkable that the contrast variations can be explained by a common dependence of the stellar line shape with latitude. In that scenario, a joint RMR fit to the ESPRESSO and HARPS-N datasets allow us to constrain the stellar inclination, thus breaking the sky-projection degeneracy and measuring the 3D spin-orbit angles of HD\,3167b (29.5$\stackrel{+7.2}{_{-9.4}}^{\circ}$) and HD\,3167c (107.7$\stackrel{+5.1}{_{-4.9}}$). 

HD\,3167b is close to orbiting within the stellar equatorial plane, while HD\,3167c orbits above the poles of the star, and the orbits of the two planets are nearly perpendicular (mutual inclination 102.3$\stackrel{+7.4}{_{-8.0}}^{\circ}$). This unusual orbital architecture likely traces different dynamical pathways for the planets. Building on the results from \citet{Dalal2019}, we propose that the evolution of HD\,3167b has always been dominated by its interactions with the star, so that its orbit retained its primordial alignment, while the evolution of HD\,3167c and d was influenced by secular interactions with an outer companion, so that their orbits were gently excited to their present high inclination. Follow-up observations of the system via velocimetry, transits, and direct imaging are needed to refine the properties of the planets, in particular their eccentricity, and to confirm the presence of the companion hinted in RV observations of the system (\citealt{Dalal2019}). Exoplanets on perpendicular orbits, resulting from gravitational interactions with outer companions, might actually be a more common occurrence than previously considered (\citealt{Albrecht2021}).  \\


\begin{acknowledgements}
We thank the referee for their appreciative review of our work. We warmly thank G. Guilluy and A. Gressier for their help with the ephemeris of HD\,3167c. The authors acknowledge the ESPRESSO project team for its effort and dedication in building the ESPRESSO instrument. This work has been carried out in the frame of the National Centre for Competence in Research ``PlanetS'' supported by the Swiss National Science Foundation (SNSF). This project has received funding from the European Research Council (ERC) under the European Union's Horizon 2020 research and innovation programme (project {\sc Four Aces}, grant agreement No 724427; project {\sc Spice Dune}, grant agreement No 947634, project {\sc SCORE}, grant agreement No 851555). V.A. acknowledges the support from FCT through Investigador FCT contracts nr. IF/00650/2015/CP1273/CT0001. This work was supported by FCT - Funda\c{c}\~{a}o para a Ci\^encias e a Tecnologia through national funds and by FEDER through COMPETE2020 - Programa Operacional Competitividade e Internacionaliza\c{c}\~{a}o by these grants: UID/FIS/04434/2019; UIDB/04434/2020; UIDP/04434/2020; PTDC/FIS-AST/32113/2017 \& POCI-01-0145-FEDER-032113; PTDC/FIS-AST/28953/2017 \& POCI-01-0145-FEDER-028953; PTDC/FIS-AST/28987/2017 \& POCI-01-0145-FEDER-028987. M.R.Z.O. acknowledges funding from the Spanish Ministery of Science and Innovation through project PID2019-109522GB-C51. R. A. is a Trottier Postdoctoral Fellow and acknowledges support from the Trottier Family Foundation. This work was supported in part through a grant from FRQNT. ASM acknowledges financial support from the Spanish Ministry of Science, Innovation and Universities Innovation (MICINN) under the 2019 Juan de la Cierva Programme and the project AYA2017-86389-P. O.D.S.D. is supported in the form of work contract (DL 57/2016/CP1364/CT0004) funded by FCT. 
\end{acknowledgements}

\bibliographystyle{aa} 
\bibliography{biblio} 

\begin{appendix}

\section{Correlation diagrams for the HD\,3167\,b transit}
\label{apn:corr_cont}

\begin{figure*}
\begin{minipage}[h!]{\textwidth}
\includegraphics[trim=0cm 0cm 0cm 0cm,clip=true,width=0.6\columnwidth]{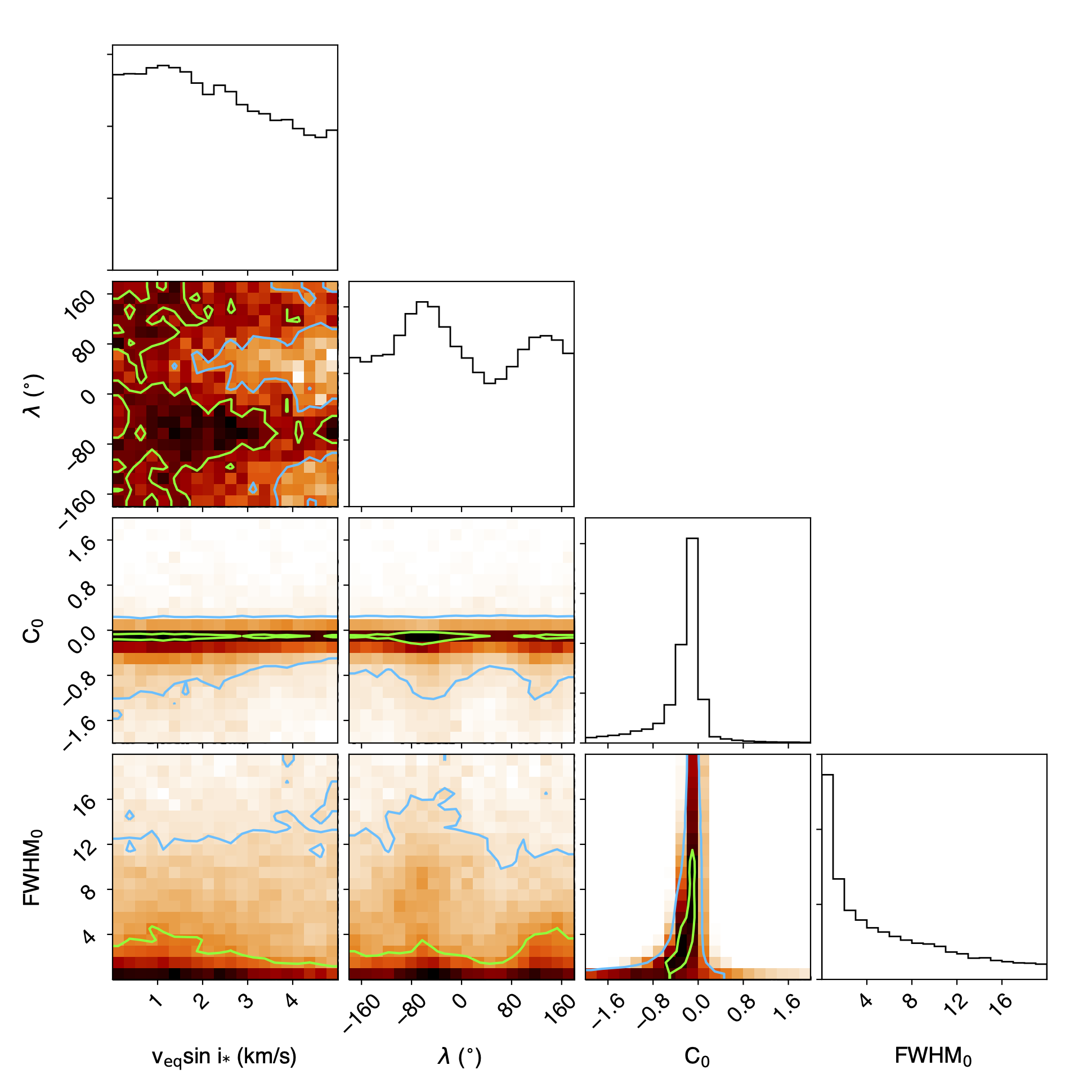}
\centering
\end{minipage}
\caption[]{Correlation diagrams for the probability distributions of the RMR model parameters, for the fit performed in the blue continuum of the CCF$_\mathrm{intr}$. }
\label{fig:PDF_HD3167b_blue}
\end{figure*}

\begin{figure*}
\begin{minipage}[h!]{\textwidth}
\includegraphics[trim=0cm 0cm 0cm 0cm,clip=true,width=0.6\columnwidth]{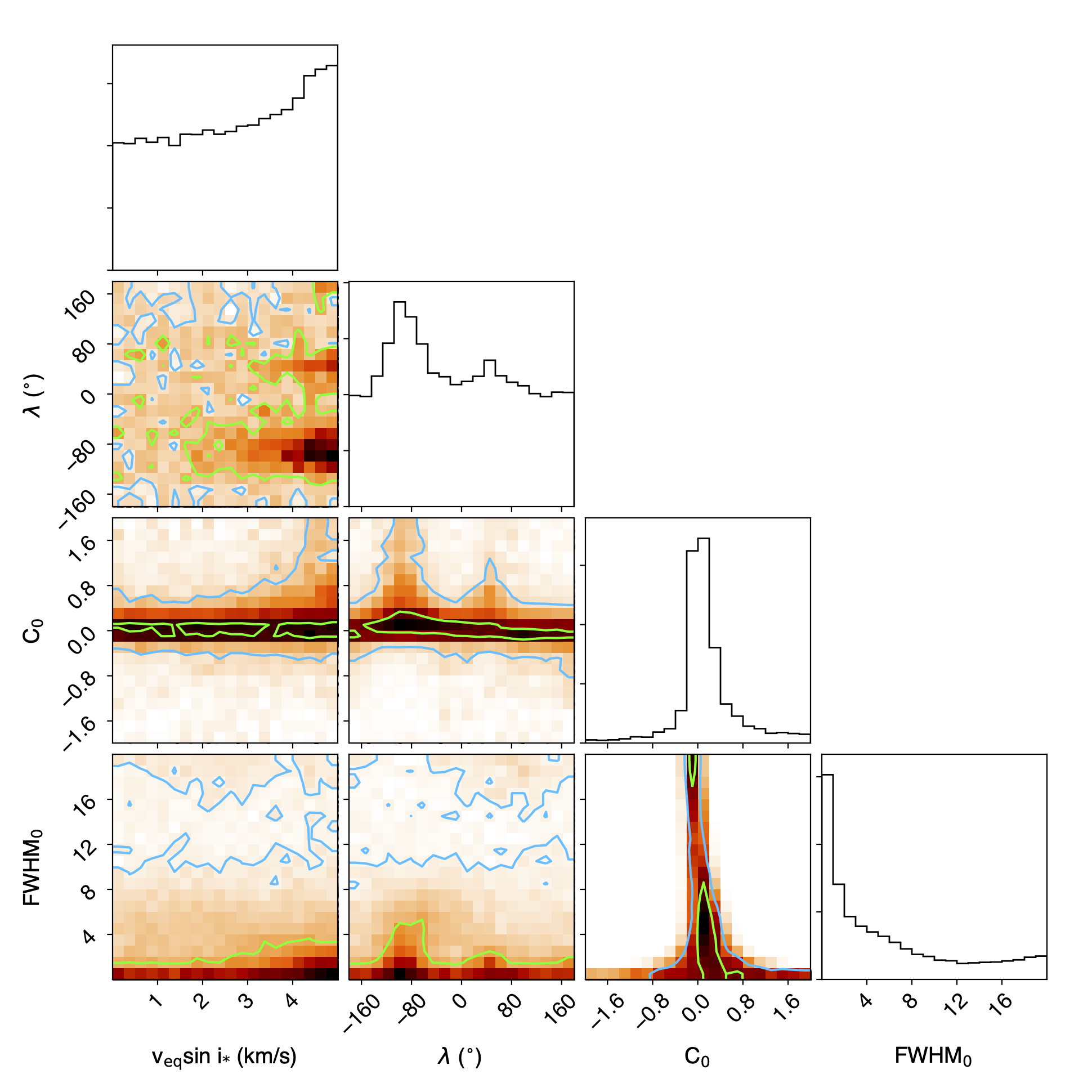}
\centering
\end{minipage}
\caption[]{Correlation diagrams for the probability distributions of the RMR model parameters, for the fit performed in the red continuum of the CCF$_\mathrm{intr}$. }
\label{fig:PDF_HD3167b_red}
\end{figure*}

\begin{figure*}
\begin{minipage}[h!]{\textwidth}
\includegraphics[trim=0cm 0cm 0cm 0cm,clip=true,width=0.6\columnwidth]{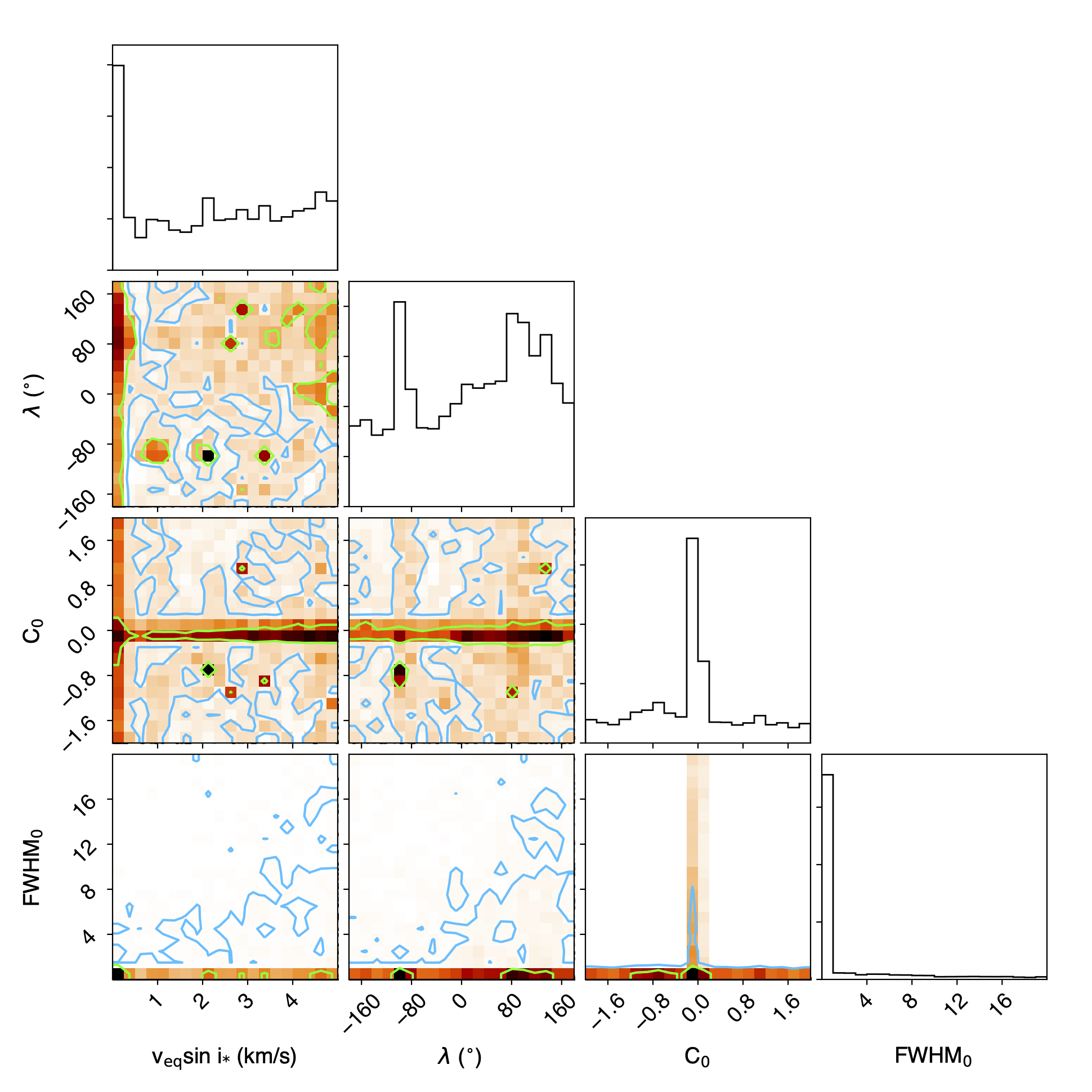}
\centering
\end{minipage}
\caption[]{Correlation diagrams for the probability distributions of the RMR model parameters, for the fit performed on the post-transit CCF$_\mathrm{loc}$.}
\label{fig:Corr_diag_postTR}
\end{figure*}

\end{appendix}

\end{document}